\documentstyle[11pt,aaspp4]{article}

\input epsf

\def\gtorder{\mathrel{\raise.3ex\hbox{$>$}\mkern-14mu
             \lower0.6ex\hbox{$\sim$}}}
\def\ltorder{\mathrel{\raise.3ex\hbox{$<$}\mkern-14mu
             \lower0.6ex\hbox{$\sim$}}}

\begin{document}

\title{Global Probes of the Impact of Baryons on Dark Matter Halos}

\author{C.S. Kochanek and Martin White }

\affil{Harvard-Smithsonian Center for Astrophysics, 60 Garden St.,
  Cambridge, MA 02138}
\affil{email: ckochanek, mwhite@cfa.harvard.edu}

\begin{abstract}
The halo mass function, $dn/dM$, predicted by hierarchical clustering models 
can be measured indirectly using dynamical probes like the distribution of
gravitational lens image separations, $dn/d\Delta\theta$, or halo circular 
velocities, $dn/d v_c$.  These dynamical variables depend on the halo 
structure as well as the halo mass.  Since baryonic physics, particularly
cooling, significantly modifies the central density structure of dark matter
halos, both observational distributions show a feature corresponding to the
mass scale below which the baryons in the halo can cool (i.e. galaxies versus
clusters).  We use simplified
but self-consistent models to show that the structural changes to the halos
produced by the cooling baryons explain both distributions.  Given a fixed
halo mass function, matching the observed image separation distribution or
local velocity function depends largely on $\Omega_b$ through its effects
on the cooling time scales.  These baryonic effects on the halo structure
also affect the evolution of the velocity function of galaxies with redshift.
\end{abstract}


\section{Introduction}

The number density, spatial distribution and properties of dark matter halos
are now well understood in models based on hierarchical clustering
(e.g.~Jenkins et al.~\cite{JFWCCEY};
 Sheth \& Tormen~\cite{SheTor};
 Navarro, Frenk \& White~\cite{NFW};
 Moore et al.~\cite{MGQSL}), 
thanks in large part to numerical experiments involving high resolution N-body
simulations.
However, the relationship between these dark matter halos and astrophysical
objects can be complicated by the modifications to the halos produced by
baryonic physics and the dependence of our search and measurement methods on
their baryonic properties.
The dominant divide in the observed properties of halos is between galaxies
and groups or clusters of galaxies.
Physically this division is between halos in which the baryons have cooled
and formed stars as compared to halos which have not
(Silk~\cite{Sil77};
 Rees \& Ostriker~\cite{ReeOst};
 White \& Rees~\cite{WhiRee};
 Blumenthal et al.~\cite{Blu84}).

In this paper we explore how the halo mass function, $dn/dM$, is globally
related to the masses or internal velocities of the objects we observe.
By using dynamical probes of the halos rather than luminosities we can
try to avoid two problems.  First, dynamical probes can provide a common
measurement scale for both galaxies and clusters, allowing us to explore
the location and properties of the boundary between the two types of 
astronomical objects.  Second, by focusing on dynamical properties of the 
halos rather than luminosities, we partly avoid the issues related to 
star formation and feedback which are central to any relationship between the
halo mass function and the luminosity function, $dn/dL$.

We cannot, however, simply ignore the effects of the baryons, because the
baryons significantly modify the halo structure when they cool
(e.g Blumenthal et al.~\cite{Blum86};
 Mo, Mao \& White~\cite{MMW};
 Cole et al.~\cite{Cole00};
 Gonzalez et al.~\cite{GWBKP}).
Both massive galaxies and small groups of galaxies can have internal velocity
dispersions $\sim 200$~km~s$^{-1}$, but it is difficult to infer from that
fact the total, and presumably different, masses of the two systems.
Thus, any global relationship between the halo mass function, $dn/dM$, and
the observed properties of galaxies, groups and clusters should show a feature
at the cooling mass scale $M_c$ dividing cooled and uncooled halos.
In this paper we quantitatively develop this theme by combining
Press-Schechter~(\cite{PreSch}) based estimates of the mass function with
adiabatic compression models (e.g.~Blumenthal et al.~\cite{Blum86})
to estimate the effects of the cooling baryons on the structure of the halo.
This allows us to make relatively self-consistent estimates for the effects of
the baryons in distorting the initial halo mass function into what our
observational probes can measure.

Many of the issues we will explore have been treated in earlier studies.
That differences in cooling time scales create the division between
galaxies and clusters is well known, and most of the effects we discuss
are explicitly included in semi-analytic models\footnote{Since the cooling
models adopted in the current generation of semi-analytic models are very
similar, we will collectively refer to them as ``SA'' hereafter} of galaxy
formation
(e.g. Lacey \& Silk~\cite{LacSil};
 White \& Frenk~\cite{WhiFre};
 Cole et al.~\cite{CAFNZ94};
 Baugh, Cole \& Frenk~\cite{BauColFre};
 Kauffman, White \& Guiderdoni~\cite{KauWhiGui};
 Kauffman et al.~\cite{KCDW};
 Somerville \& Primack~\cite{SomPri};
 Benson et al.~\cite{BCFBL};
 Cole et al.~\cite{Cole00}).
However our focus here is somewhat different and we highlight two issues which
have not previously been emphasized.
First, we focus on global probes of the mass function on both galaxy and
cluster scales and the role of cooling in producing a measurable feature in
these observational distributions.
Second, we emphasize the importance of structural changes in the mass
distribution of the dark matter
halo rather than the effects of cooling on the star formation rate and
luminosity/colors of galaxies which are a major focus of semi-analytic work.

The outline of the paper is as follows.
In \S\ref{sec:halo} we review the modified Press-Schechter formalism which we
use for determining the mass function of halos, adiabatic compression 
models for the effects of the cooling baryons on the structure of the halos,
and simple cooling models for the determining the cooling mass scale.
In \S\ref{sec:lens} we use these models to explore the distribution of
image separations in gravitational lenses, and in \S\ref{sec:velocity} 
we illustrate the effects of the cooling baryons on the distribution of
halos in dynamical velocities.
In \S\ref{sec:conclusions} we summarize our results.

\section{The Properties of Halos} \label{sec:halo}

Our models have three elements.
First, we need to estimate the mass function of halos, $dn/dM$, and
the formation epoch of each halo.
Second, we need to estimate how the structure of the halo is modified by
the cooling of the baryons.
Third, we need to estimate whether the baryons in the halo have cooled.
We will use a fixed cosmological model throughout the paper, the so-called
``concordance'' model of Ostriker \& Steinhardt~(\cite{OstSte}) which
provides an acceptable fit to a wide range of current observations.
This is a $\Lambda$CDM model with a matter density (in units of the critical
density) $\Omega_m=0.3$, a Hubble constant $H_0=67$~km~s$^{-1}$~Mpc$^{-1}$,
and a baryon density $\Omega_b=0.04$ normalized to match the abundance of rich
clusters of galaxies ($\sigma_8=0.9$).

\subsection{The Halo Mass Function and Formation Time} \label{sec:PS}

To calculate the cosmological mass function of dark matter halos we use a
fit to the results of the Virgo simulations
(Sheth \& Tormen~\cite{SheTor}; Jenkins et al.~\cite{JFWCCEY}).
We assign halos of mass $M$ to peaks of height $\nu$
\begin{equation}
  \nu \equiv \left( {\delta_c\over \sigma(M)} \right)^2
\end{equation}
where $\delta_c=1.69$ and $\sigma(M)$ is the rms fluctuation in the matter
density smoothed with a top-hat filter on a scale $R^3=3M/4\pi \bar{\rho}$
with $\bar{\rho}$ the mean matter density.
Press-Schechter~(\cite{PreSch}) theory postulates that the mass function can
be cast in terms of $\nu$ into a universal form
\begin{equation}
  {M\over\bar{\rho}} {dn\over dM}dM = f(\nu)d\nu
\end{equation}
where $f(\nu)$ is known as the multiplicity function.
We use\footnote{We use this form rather than the fit from
Jenkins et al.~(\protect\cite{JFWCCEY}) because the range of validity
of the latter is too small for our purposes.}
(Sheth \& Tormen~\cite{SheTor})
\begin{equation}
  \nu f(\nu) = A(1+\nu'^{-p}) \nu'^{1/2} e^{-\nu'/2}
\label{eqn:fnu}
\end{equation}
where $p=0.3$ and $\nu'=0.707\nu$.  The normalization constant $A$ is fixed
by the requirement that all of the mass lie in a given halo
\begin{equation}
  \int f(\nu) d\nu = 1 \qquad .
\end{equation}
The Press-Schechter form is given by $p\to 0$ and $\nu'\to\nu$.

Later we will need to know the distribution of formation times for our
dark matter halos.
Following Kitayama \& Suto~(\cite{KitSut}) and Newman \& Davis~(\cite{NewDav})
we model the formation time distribution using the extended Press-Schechter
theory outlined in Lacey \& Cole~(\cite{LacCol}).
The formation time distribution, $dp/dt_{\rm form}$, is an integral
over mass of the progenitors which merge to form a given halo at the
observed redshift (Lacey \& Cole~\cite{LacCol}; Eq. 2.19).
Because this integral can be difficult to evaluate numerically we in fact
integrate the cumulative probability distribution and obtain the desired
differential distribution by finite difference.
Since all of the time dependence is in the threshold parameters $\delta(t)$
this can be cast into an elegant form and is very stable numerically.

\subsection{Adiabatic Compression} \label{sec:adiabatic}

The dark matter halo number densities calculated above represent
``primordial'' halos.  To make contact with observations it is necessary to
take into account the effects of baryons on halo structure and dynamics.
We follow Mo et al.~(\cite{MMW}) in estimating the modifications in the mass
distribution of the halos created by the cooling baryons, though similar
approaches are used by Dalcanton, Spergel \& Summers~(\cite{DalSpeSum}),
Cole et al.~(\cite{Cole00}) and Gonzalez et al.~(\cite{GWBKP}).

We assume that the halos all have spherical profiles depending only on the
mass and we neglect any substructure or halos-within-halos.
Specifically we model the profile with the ``universal'' form\footnote{Our
results will not depend sensitively on this choice.  The alternate form
of Moore et al.~(\protect\cite{MGQSL}) would work just as well.}
described by Navarro, Frenk \& White~(\cite{NFW}; hereafter NFW)
\begin{equation}
  \rho(r) \propto {1\over x(1+x)^2}
\label{eqn:nfw}
\end{equation}
where $x=r/r_s$ is the radius measured in units of a characteristic scale
$r_s$.  Each halo can then be characterized by two numbers.  Rather than $r_s$
and the constant of proportionality in Eq.~(\ref{eqn:nfw}), we take these to
be the virial mass and the concentration.
The virial mass $M_{\rm vir}$ is the total mass inside the virial radius
$r_{\rm vir}$: the radius within which the mean density exceeds the critical
density by a factor of $\Delta_c(z)$.
In an Einstein-de Sitter model $\Delta_{\rm c}=18\pi^2\simeq 178$.
It is lower for $\Omega_m<1$, taking the value $\Delta_{\rm c}\simeq 100$ for
the ``concordance'' cosmology.
We estimated the concentration of the halos,
\begin{equation}
   c \equiv {r_{\rm vir} \over r_s } = { 9 \over 1 + z} 
     \left( { M_{\rm vir} \over 8.12 \times 10^{12} M_\odot} \right)^{-0.14},
\end{equation}
using the average relation found by Bullock et al.~(\cite{BKSSKKPD};
see also Eke, Navarro \& Steinmetz~\cite{EkeNavSte}).
Physically this suggests that $r_s$ is roughly constant with redshift.
Finally, the halo is assumed to have angular momentum $J$ specified by its
spin parameter $\lambda = J|E|^{1/2}/GM_{\rm vir}^{5/2}$, where the binding
energy $|E|$ is computed using the virial theorem.  
    
We model the disk of the galaxy
as an exponential disk characterized by
mass $M_d=m_d M_{\rm vir}$ and scale length $r_d$.  The disk is assumed to
have angular momentum $J_d = j_d J$, and this is used to determine the disk
scale length.
If $j_d = m_d$, the specific angular momentum of the disk is the same as that
of the halo.
Unlike Mo et al.~(\cite{MMW}), who considered adding a bulge component as a
point mass, we added a bulge modeled as a Hernquist~(\cite{Her}) profile with
mass $M_b = m_b M_{\rm vir}$ and a Hernquist scale length $a_b$.
In contrast to the disk, there is no conserved quantity like the angular
momentum that can be used to determine the bulge scale length.
Instead we simply used a phenomenological scaling\footnote{In the photometric
survey of spiral galaxies by de Jong~(\cite{deJ}), bulge effective radii are
typically $R_e\simeq r_d/10$ of the disk scale length $r_d$, with a
logarithmic scatter of $0.17$~dex. For a Hernquist model, the scale length is
$a=0.45R_e$ of a de Vaucouleurs effective radius $R_e$, so we use
$a=0.045r_d$.} that $a=0.045 r_d$.
We assumed that the total specific angular momentum of the baryons was
the same as the dark matter, but that all of the angular momentum is in
the disk component ($j_d=m_d+m_b$) while the bulge has none.

The dark matter is adiabatically compressed by the cooled baryons, which
for a spherical system of particles on circular orbits means conserving
the mass and angular momentum (e.g.~Blumenthal et al.~\cite{Blum86}).
Thus, a dark matter particle initially orbiting at radius $r_i$ ends up at
radius $r$ where $r_i M_i(r_i)=r M_f(r)$ and $M_i(r_i)$ and $M_f(r)$ are
the initial and final mass distributions.
The initial mass distribution is simply that of the NFW halo, $M_{NFW}(r_i)$,
while the final mass distribution is
\begin{equation}
    M_f(r) = M_d(r) + M_b(r) + (1-m_d-m_b) M_{NFW} (r_i). 
\label{eqn:adcomp}
\end{equation}
After making an initial estimate for the disk scale length, $M_f(r)$ is
found by iteratively solving Eq.~(\ref{eqn:adcomp}) while adjusting the
disk and bulge scale lengths to satisfy the disk angular momentum constraint 
(see Mo et al.~\cite{MMW})
\begin{equation}
  J_d = j_d J = 2\pi \int_0^{r_{\rm vir}} R\, dR\ v_c(R) R\, \Sigma(R) 
\end{equation}
where $v_c(R)$ is the rotation curve in the disk and $\Sigma(R)$ is the
surface mass density of the disk.  The rotation curve $v_c(R)$ is computed 
assuming that the bulge and the halo remain spherical while the disk is an 
infinitely thin exponential disk.

\subsection{A Very Simple Cooling Model} \label{sec:cooling} 

We treat cooling using two very simple models.
The first is simply adding a cooling mass scale $M_c$ by hand.
We assume that the probability that a halo has cooled is given by
\begin{equation}
  P_{\rm cool}(M) = 1 -
  \left[ 1 + \exp\left({ \log M -\log M_c)\over \Delta M_c }
  \right)\right]^{-1}
\label{eqn:pcool}
\end{equation}
with a cooling mass scale $M_c$ and an (arbitrary!) logarithmic distribution
width of $\Delta M_c=0.1$ to smooth the transition.  Half of the halos with
$M=M_c$ have cooled, and the total distribution of halos is the sum of the
cooled and uncooled distributions.  For any observational test we can vary
the cooling mass scale $M_c$ until we achieve a reasonable match.

The second model is based on a simplified version of the cooling model used by
Cole et al.~(\cite{Cole00}) in their semi-analytic models of galaxy formation.
We will use this model both to check whether our estimate of the cooling
mass scale (\S\ref{sec:coolscale}) is reasonable and as a complete
self-consistent model (\S\ref{sec:coolmodel}).
As we are interested only in the boundary between galaxies and clusters, we
neglect the complicated details of star formation and feedback and consider
only cooled mass fractions.
We assume that when the halo collapses, the gas is heated to the virial
temperature $k_BT_{\rm vir} = {1\over 2}\mu m_p v_{\rm vir}^2$ where $m_p$ is
the proton mass and $\mu$ is the mean molecular weight.
We assume that the initial density distribution of the gas is
$\rho_{\rm gas} \propto (r^2+r_c^2)^{-1}$ for $r<r_{\rm vir}$
where $r_c=r_s/3$.
This is the phenomenological model which Cole et al.~(\cite{Cole00}) develop
based on more realistic collapse simulations with hydrodynamics.
Given the temperature, the gas density and the
Sutherland \& Dopita~(\cite{Suth93})
cooling curves, we can estimate the cooling time as a function of radius,
\begin{equation}
   \tau_{\rm cool}(r) =
   {3\mu^2 m_p^2 v_{\rm vir}^2\over 4\rho_{\rm gas}
   \Lambda_N(T_{\rm vir})}
\end{equation}
where we have fixed the metallicity to $Z=Z_\odot/3$ for simplicity.

As a consistency check, we can estimate the age of a halo and the time for
a given fraction of the baryons to cool to see if they match at a mass scale
comparable to our estimated cooling mass scale $M_c$ (\S\ref{sec:coolscale}).
We can also use the cooling model to determine the cooled baryon fraction of 
each halo.
Given the average age, $t_{\rm form}(M,z)$, for halos of mass $M$ and redshift
$z$, we calculate the cooled baryonic mass fraction $f_{\rm cool}(M,z)$ by the
mass fraction inside the radius where the cooling time equals the age,
$t(z)-t_{\rm form}(M,z)=\tau_{\rm cool}(r_{\rm cool})$.
If the global baryon fraction is  $(m_d+m_b)_0$, then we model the halo with
an adiabatic compression model having a baryon fraction
$m_d+m_b=f_{\rm cool}(M,z)(m_d+m_b)_0$, eliminating the cooling mass scale
$M_c$ as a parameter.
Assuming that all halos start as fair samples of the universe, the global
baryon fraction $(m_d+m_b)_0=\Omega_b/\Omega_m$.
Naively then, the only parameter of the model is the baryon density $\Omega_b$.
In practice however, star formation complicates the interpretation.
While $\Omega_b$ controls the initial cooling of the halo, star formation can
reheat the baryons once they have cooled so that the cool baryon fraction in
a galaxy is less than the fraction which cooled initially (see e.g.~SA).
Since the adiabatic compression is due only to the cold baryons, we will call
our parameter $\Omega_{b,{\rm cool}}$ and we should find that
$\Omega_{b,{\rm cool}} \leq \Omega_b$.

\section{The Distribution of Gravitational Lens Image Separations }
\label{sec:lens}

The distribution of image separations $\Delta\theta$ in gravitational lenses
is directly related to the redshift-averaged distribution of halos $dn/dM$,
where the conversion depends on the halo mass, redshift, and internal
structure.  Given a large well-defined lens sample, the image separation
distribution will be the cleanest global probe of the halo mass function
because the selection of the lenses is independent of the flux or surface 
brightness of the baryons in the halo.  Lens surveys also avoid the
distinction between low-mass (galaxy) and high-mass (cluster) systems 
required for any other global estimate of the mass function.
Wide separation lenses (usually defined by $\Delta\theta \geq 3\farcs0$) are
created  by groups and clusters of galaxies and small separation systems are
created by individual galaxies.
We will show, however, that a successful model of the separation distribution
must correctly include both the mass scale below which the baryons
significantly modify the structure of the halo and the changes in the
structure of the halos produced by the cooled baryons.  

We will examine the distribution of image separations of the lenses found in
the Cosmic Lens All-Sky Survey (CLASS; e.g.~Browne \& Myers~\cite{Bro00}) for
lensed flat-spectrum radio sources.  The primary CLASS survey has a nearly
uniform selection function for separations from
$0\farcs3\leq\Delta\theta\leq 6\farcs0$, with extensions to wider separations
of $6\farcs0\leq\Delta\theta\leq 15\farcs0$ (Phillips et al.~\cite{Phil00}).
These surveys have found 18 lenses, all with separations
$\Delta\theta < 3\farcs0$.
We also considered the separation distribution of all radio-selected lenses.
This sample of 27 lenses is inhomogeneously selected, but includes two wider
separation lenses (MG~2016+112 and Q~0957+561) and illustrates the
uncertainties in the tail of the distribution.  We include the angular 
selection function for small separation lenses, but with an outer limit
of $15\farcs0$ we will be able to ignore the selection function for large
separation lenses.  

\subsection{The Need for Baryons} \label{sec:baryons}

The need for baryonic processing to explain the image separation
distribution is most
obvious when we consider the previous attempts to compute the distribution
of lens separations based only on the properties of the dark matter
(e.g.~Narayan \& White~\cite{Nar88};
Kochanek~\cite{Koc95}; Wambsganss et al.~\cite{Wamb95};
Wambsganss, Cen \& Ostriker~\cite{Wamb98}; Maoz et al.~\cite{Maoz97};
Keeton~\cite{Kee98}; Mortlock \& Webster~\cite{Mortlock00}; Li \& Ostriker~\cite{LiOst},
Keeton \& Madau~\cite{Kee00}, Wyithe, Turner \& Spergel~\cite{Wy00}).
In these models, most of which were intended only to explain the wide
separation lenses, the mass function of the parent dark matter halos is
calculated using the Press-Schechter~(\cite{PreSch}) theory and its
extensions.  Given a mass function, the lens properties are then
calculated by assuming a model for the density distribution of the halos.
When the models are normalized so that they predict the correct
local abundance of massive clusters, they correctly predict that
wide separation lenses are rare.  However, they catastrophically
fail to explain the distribution of smaller separation lenses.

A purely phenomenological approach based on the local properties of galaxies,
by contrast, predicts the observed properties of lenses quite well
(e.g.~Kochanek~\cite{Koc96}, Keeton, Kochanek \& Falco~\cite{Kee98b}).
These models usually combine local galaxy luminosity functions with local
kinematic relations to predict the distribution of lenses assuming a constant 
comoving density of galaxies, although a few
studies have considered the effects of number evolution and merging
(e.g.~Mao \& Kochanek~\cite{Mao94}; Rix et al.~\cite{Rix94}).
These models have modest difficulty explaining the largest separation lenses
observed in systematic surveys (separations of $\Delta\theta\simeq 6\farcs0$),
and cannot explain the lensing effects of rich clusters.

Keeton (\cite{Kee98}) pointed out that the origin of the problem lay in
ignoring the extra physics which makes the density structure of the lenses
depend strongly on the mass scale.  Any model based on the mass function of
dark matter halos which assumes that the density distributions of the
halos vary smoothly and continuously with mass leads to predictions for
the separation distribution that include far too many wide separation
lenses compared to small separation lenses.
Keeton (\cite{Kee98}) demonstrated the effect by showing that using either
SIS (singular isothermal spheres) or NFW density profiles for all lenses
could not explain the observations.  Only by introducing a baryonic cooling
mass scale $M_c$ below which the halos were modeled as SIS lenses and above
which they were modeled as NFW halos could the observed properties be
explained.
In particular this change in structure explains why many lenses found in
groups of galaxies were associated with the galaxies in the group rather
than the group halo, even though the group halo had to be more massive than
its component galaxies.

Porciani \& Madau (\cite{PorMad}) used this hypothesis in their models for
the separation distribution of lenses and found that the mass scale below
which the halos had to cool and have the SIS density profiles was
$M_c=3.5\times 10^{13} M_\odot$.
We shall find a similar cooling mass scale, though our model is slightly
different than that of Keeton~(\cite{Kee98}) and
Porciani \& Madau (\cite{PorMad}).
We also show that such a mass scale is naturally predicted by the
properties of halos in hierarchical structure formation models if the
baryon density is near the value preferred by big-bang nucleosynthesis
(e.g.~O'Meara et al.~\cite{OMea}).

\begin{figure}
\epsscale{0.25}
\plottwo{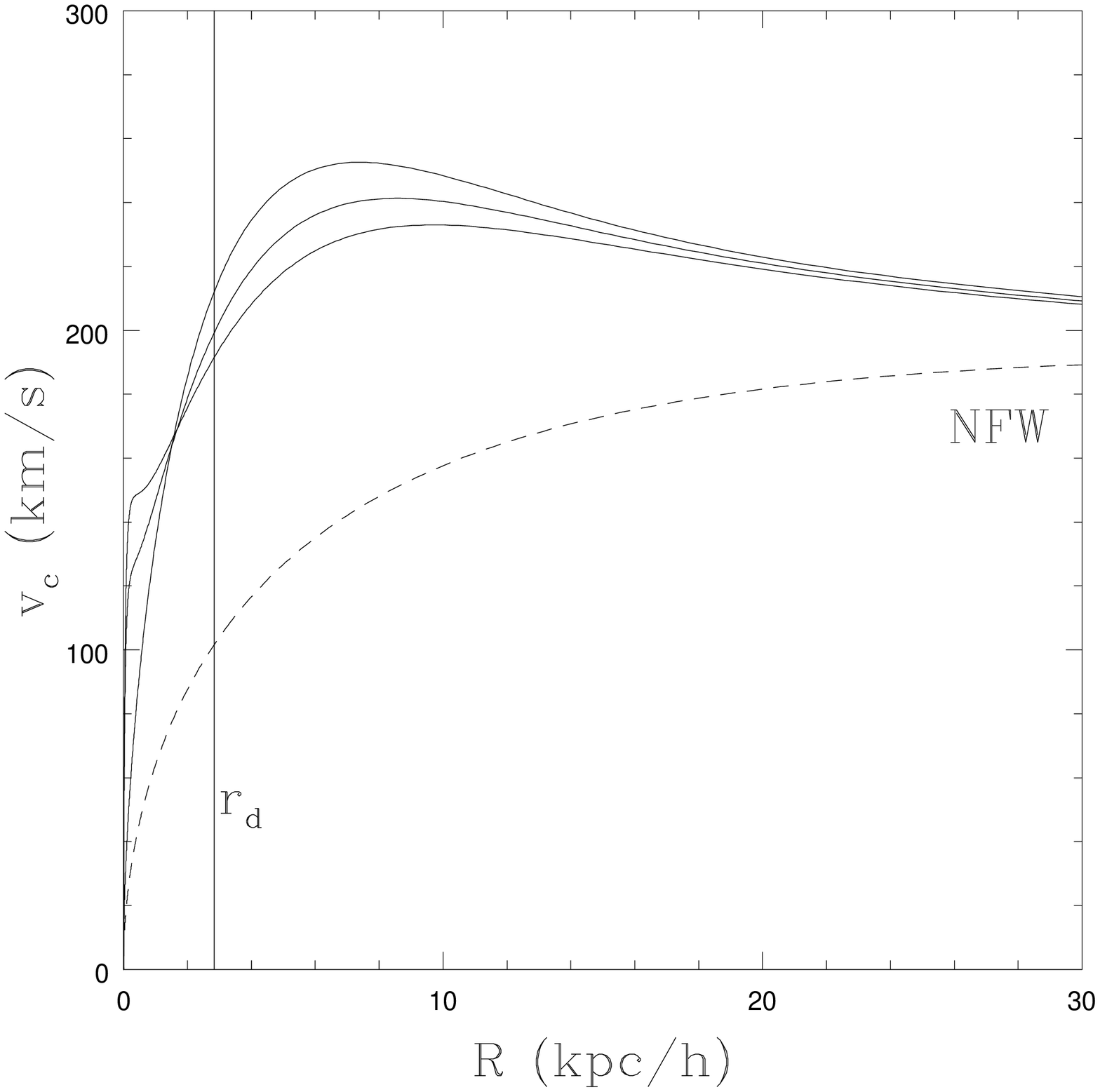}{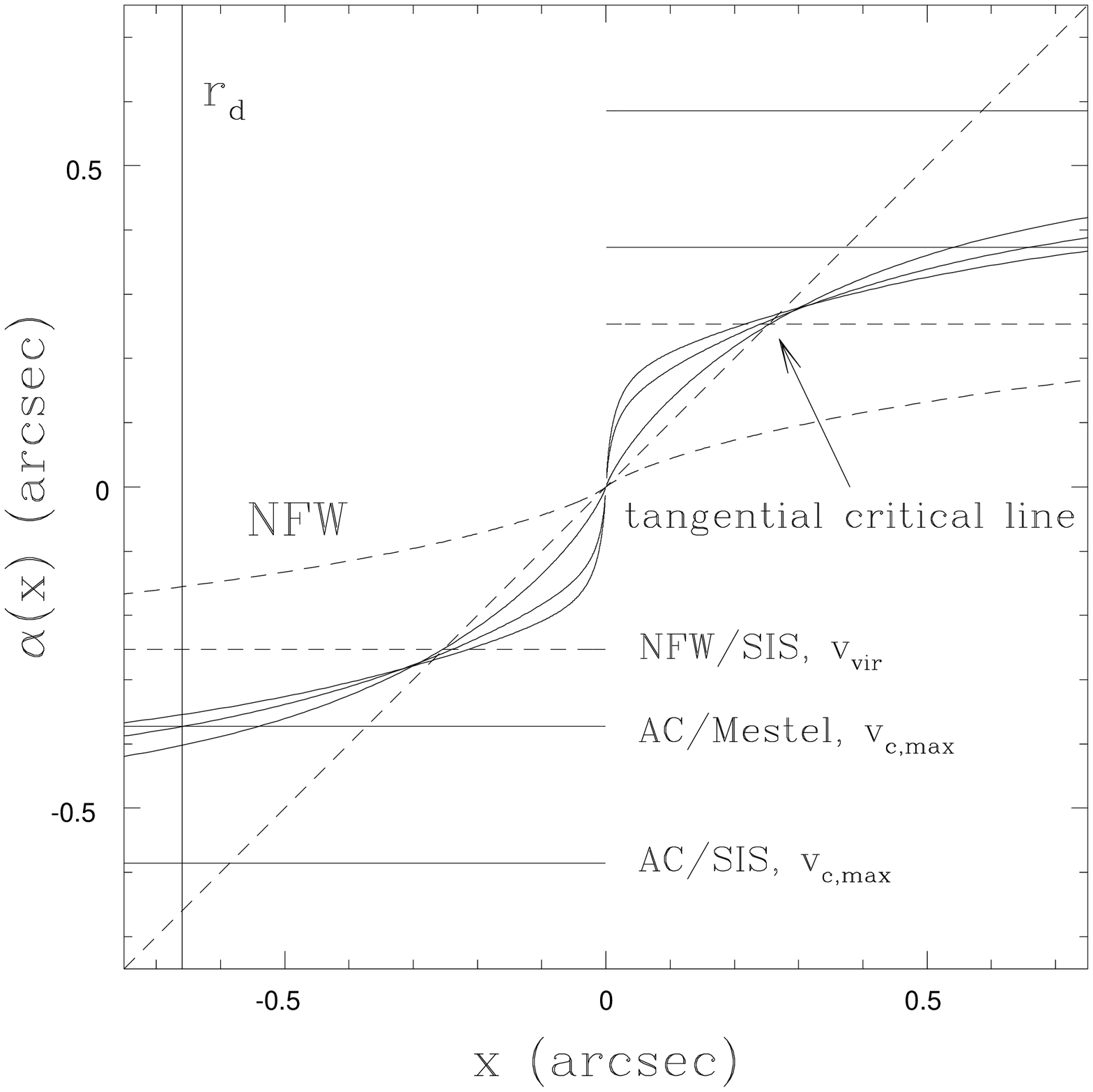}
\caption{(Left) The rotation curves for $10^{12}M_\odot$ halos at $z_l=0.5$
with concentration $c=8$.
The dashed curve shows the rotation curve of the initial NFW halo.
The solid curves show the rotation curves found after the compression of
the halo by the baryons assuming a baryonic mass fraction of $m_d+m_b=0.05$,
a spin parameter $\lambda=0.04$, and that the disk contains all the initial
baryonic angular momentum, $j_d=0.05$.
The three solid curves are for bulge-to-disk mass ratios $m_b/m_d=0$ (highest peak $v_c$),
$0.1$ and $0.2$ (lowest peak $v_c$) respectively.  The vertical line shows
the disk scale length.
(Right) The bending angles, $\alpha(x)$, produced by the same halos assuming
a source redshift of $z_s=2.0$.  The dashed curve shows the deflection produced 
by the initial NFW halo.  The solid curves show the bending angles produced after 
the compression of the halo by the baryons where the central deflection profile
becomes steeper as we increase the bulge fraction. For comparison, the horizontal 
dashed lines show three simple comparison models:  an SIS lens normalized by the
halo virial velocity $v_{vir}$ (NFW/SIS, $v_{vir}$), an SIS lens normalized by 
the peak circular velocity of the compressed halo, (AC/SIS, $v_{\rm{c,max}}$),  
and a Mestel disk lens normalized by the peak circular velocity of the compressed 
halo, (AC/Mestel, $v_{\rm{c,max}}$). The tangential critical line $x_+$ of a lens 
(the Einstein ring) is located at the point where the (dashed) $45^\circ$ line 
intersects the bending angle, and the radial critical line $x_-$ is located where 
a $45^\circ$ line is tangent to the bending angle. An arrow points to the location 
of the tangential critical line of the adiabatically compressed models. The
vertical solid line shows the disk scale length. }
\label{fig:bend}
\end{figure}

\subsection{The Qualitative Effects of Baryons on Lensing Properties}
\label{sec:cool}

We start by illustrating the effects of the baryons on the lensing properties
of a $M_{\rm vir}=10^{12}M_\odot$ halo at redshift $z_l=0.5$ (implying a 
concentration of $c=8$), spin parameter
$\lambda=0.04$, baryonic mass fraction $m_d+m_b=0.05$, and baryonic angular
momentum fraction $j_d=0.05$.
Fig.~\ref{fig:bend} shows the rotation curves for the initial dark matter
distribution and the final matter distribution given bulge-to-disk mass ratios
of $m_b/m_d=0.0$, $0.10$ and $0.20$.
We see the familiar boost in the circular velocity due to the disk and the
compression of the dark matter (e.g.~SA).
The bulge supports the central rotation velocity and slightly reduces the
outer rotation velocity because the angular momentum per unit mass in the
disk is rising as we increase the mass of the bulge.
It is instructive to note that the mass profile, angular momentum profile and
rotation curve of our compressed halos are quite different from what a pure
dark matter simulation would predict.
In particular the rotation curve only ``recovers'' to the pure dark matter
form at very large radius (not shown on Fig.~\ref{fig:bend}).

In Fig.~\ref{fig:bend} we also show the deflection or bending angle profile
$\alpha(x)$ produced by the lens assuming it lies at $z_l=0.5$ and the
source lies at $z_s=2$.  To simplify the problem we consider only the face-on 
lensing properties, but we should keep in mind that systems with flat disks 
are more efficient lenses when inclined (see Keeton \& Kochanek~\cite{KeeKoc}).
With these parameters, the initial dark matter distribution is a sub-critical
lens and can produce no multiple images of the source.  Adding the disk and 
compressing the dark matter pushes the lens above critical, although not by 
a large amount.  The radial and tangential critical radii are comparable, and 
the lens will generally produce visible central (odd) images which are almost 
never seen in real systems.

The problem is that exponential disks have little density contrast between
their centers and the disk scale length ($\simeq 4h^{-1}$~kpc) while an
efficient lens requires a large density contrast over this region (for
a recent discussion, see Rusin \& Ma~\cite{Rus01}).
The sensitivity of the lensing properties to the central density means that
adding even a small bulge enormously increases the efficiency of the galaxy
as a lens.
For bulge-to-disk mass ratios of $m_b/m_d=0.05$, $0.10$, and $0.20$ the
multiple imaging cross section increases by a factors of $6$, $12$ and $23$
compared to the model with no bulge.  Moreover, the models with bulges will 
generally have demagnified central images, as observed in real lenses.
In short, the details of the central density distribution are critical to the
statistical properties of the lens, and our adiabatic compression models are
inadequate for providing detailed predictions of the central density 
distribution.

The adiabatically compressed models are complicated, so it is
of some interest to see how they compare to the far simpler SIS and
face-on Mestel disk lens models.  Both models have flat rotation curves
and constant deflections equal to half the image separations, but if
normalized 
to the same circular velocity they produce different image separations of 
$\Delta\theta_{SIS} = 4\pi (v_c/c)^2 D_{LS}/D_{OS}$ and 
$\Delta\theta_{Mestel}=8 (v_c/c)^2 D_{LS}/D_{OS}$ because a thin
disk requires less mass than a spherical distribution to produce the same
circular velocity (see Keeton \& Kochanek~\cite{Kee98c}).
Because the inner rotation curves of the adiabatically
compressed models are dominated by the disks, a Mestel model normalized
to the peak circular velocity provides a better match to the deflections
of the adiabatically compressed models than an SIS model.  Using an SIS
model with a circular velocity equal to the initial halo virial velocity, 
$v_{vir}=(GM_{vir}/R_{vir})^{1/2}$, also
provides a good match to the deflection scale for this model. 
Since most of the known lenses are produced by bulge-dominated, early-type 
galaxies (see Keeton et al.~\cite{Kee98b}) and seem to require mass
distributions very similar to the SIS model (e.g.~Cohn et al.~\cite{Cohn01}),
it is interesting to note that the shape of the deflection profile becomes
more similar to that of the SIS model as we increase the bulge mass fraction.
  
\begin{figure}
\begin{center}
\leavevmode
\epsfxsize=6in \epsfbox{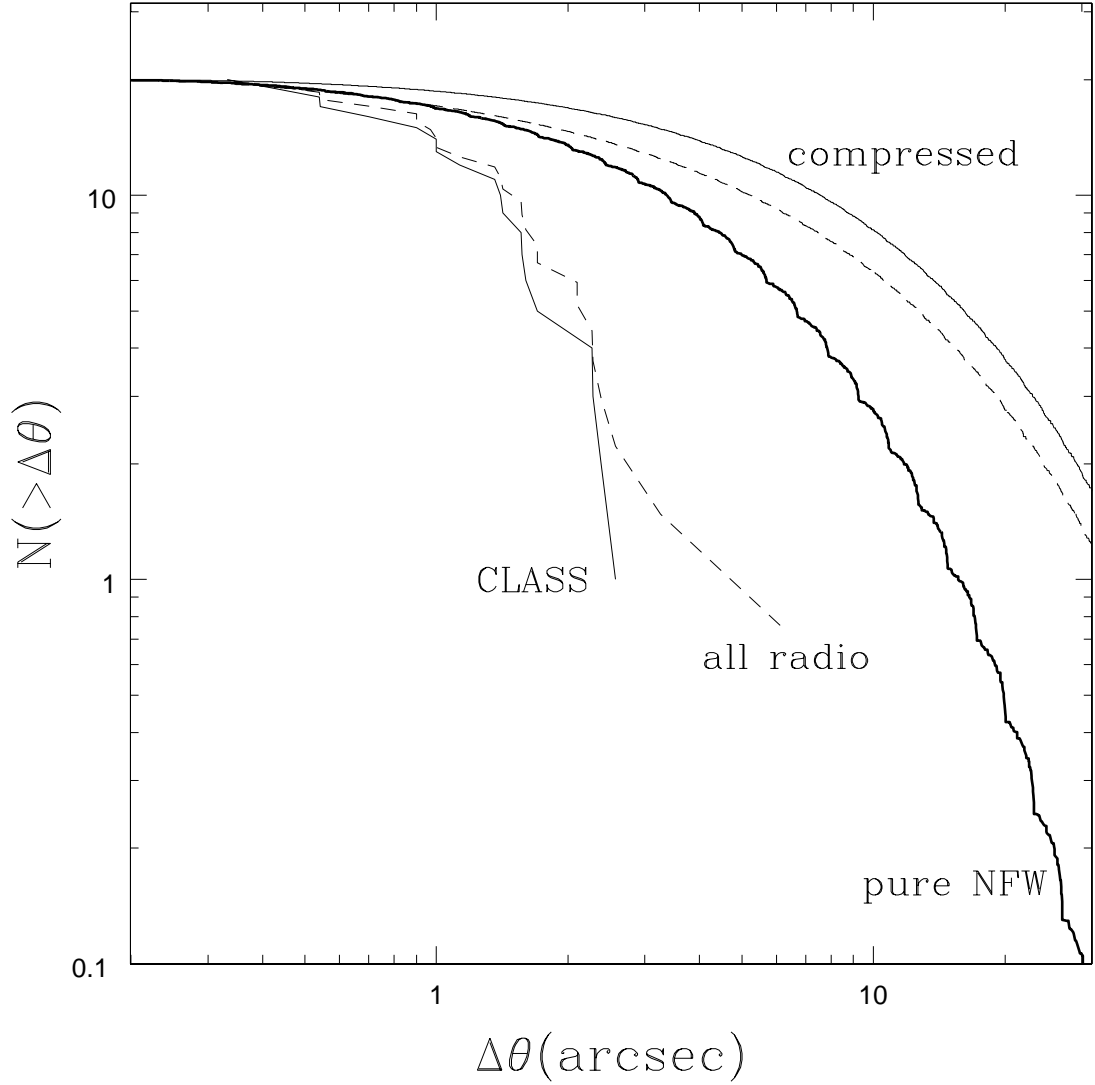}
\end{center}
\caption{Predicted separation distributions without a cooling scale.
The curves show the integral separation distribution normalized to the total
number of CLASS lenses.  The heavy solid line shows the distribution
predicted by pure NFW models while the light solid (dashed) lines shows the
distributions predicted by the adiabatic compression models with no bulge
(a 10\% baryonic mass fraction bulge). The wiggles in the pure NFW curve
are a small discretization problem. }
\label{fig:sepdist0}
\end{figure}

\subsection{The Image Separation Distribution and the Density of Baryons}
\label{sec:sep}

We explore the role of the baryons in producing the observed separation
distribution of gravitational lenses in four stages.
First, we illustrate the problem created by ignoring the baryons.
Second, we show that introducing a mass scale, $M_c$, below which the halos
are compressed explains the observed distribution.
Third, we show that the required mass scale is consistent with simple models
of cooling physics.
Fourth, we show that the fundamental variable controlling the separation
distribution is the cosmological density of cooled baryons.  
Some technical details of our calculations are relegated to
Appendix~\ref{sec:simplify}.

\subsubsection{The Cooling Scale} \label{sec:coolscale}
 
The first two stages reproduce the arguments originally developed by
Keeton (\cite{Kee98}) and Porciani \& Madau (\cite{PorMad}).  Consider,
as a standard model, a set of halos with $m_d+m_b=0.05$, $\lambda=0.04$, 
and $j_d=0.05$ with ($m_b/m_d=0.1$) and without ($m_b/m_d=0$) a bulge.
If we ignore the existence of a characteristic mass scale dividing 
galaxies from groups and clusters, then we predict image separation
distributions with far more wide separation lenses than are actually
observed.  The problem does not depend on the choice of model. 
Keeton (\cite{Kee98}) demonstrated it using NFW and SIS models for
the lenses and we illustrate it in Fig.~\ref{fig:sepdist0} for NFW
models and adiabatically compressed models both with and without 
a central bulge.  While the low mass halos have higher NFW concentration
parameters, which makes them more efficient lenses, the differences
are not large enough to produce a distribution cutting off sharply
at $\Delta\theta\simeq3\farcs0$.  

\begin{figure}
\begin{center}
\leavevmode
\centerline{\epsfxsize=3.5in\epsfbox{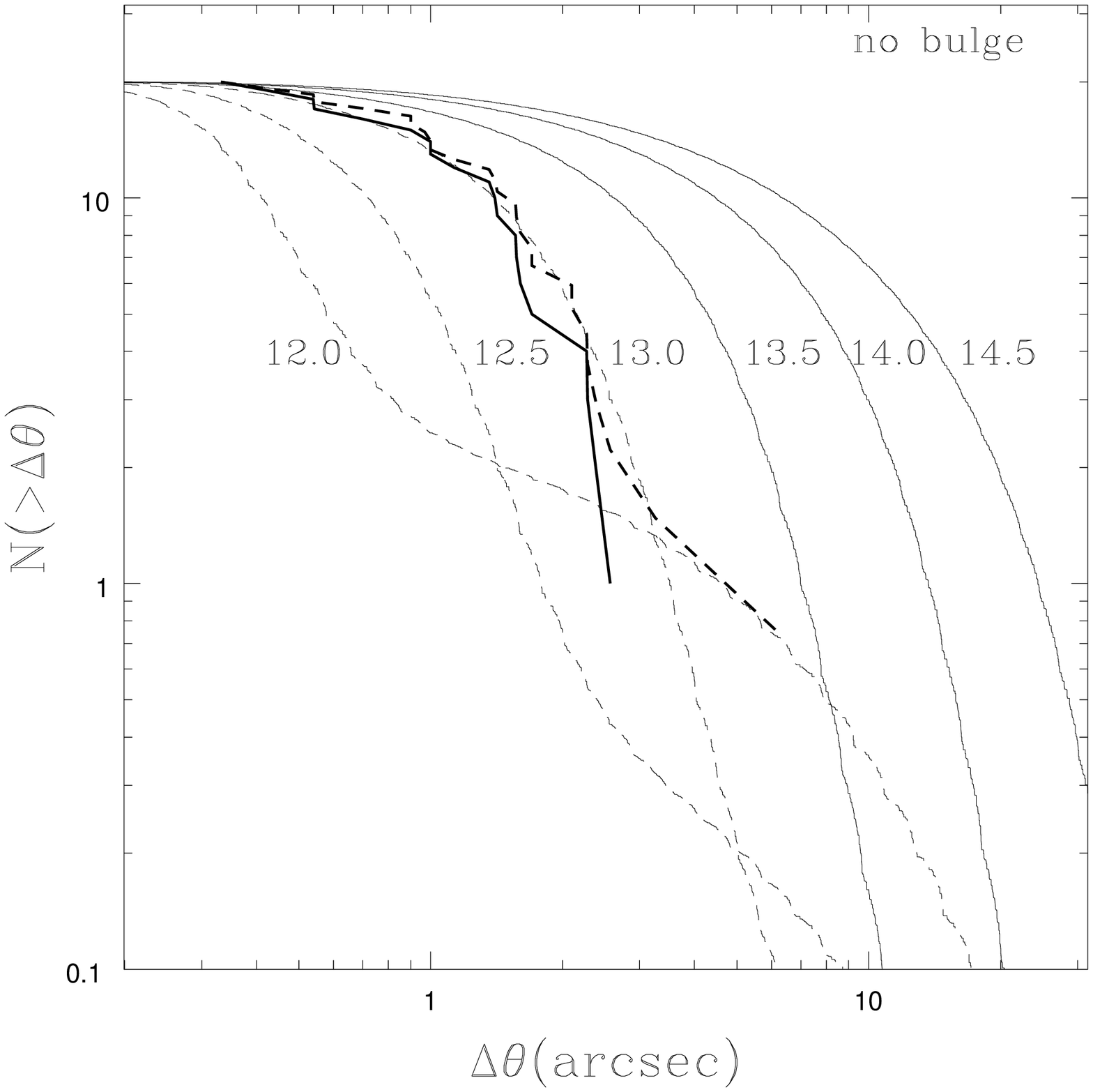}\epsfxsize=3.5in\epsfbox{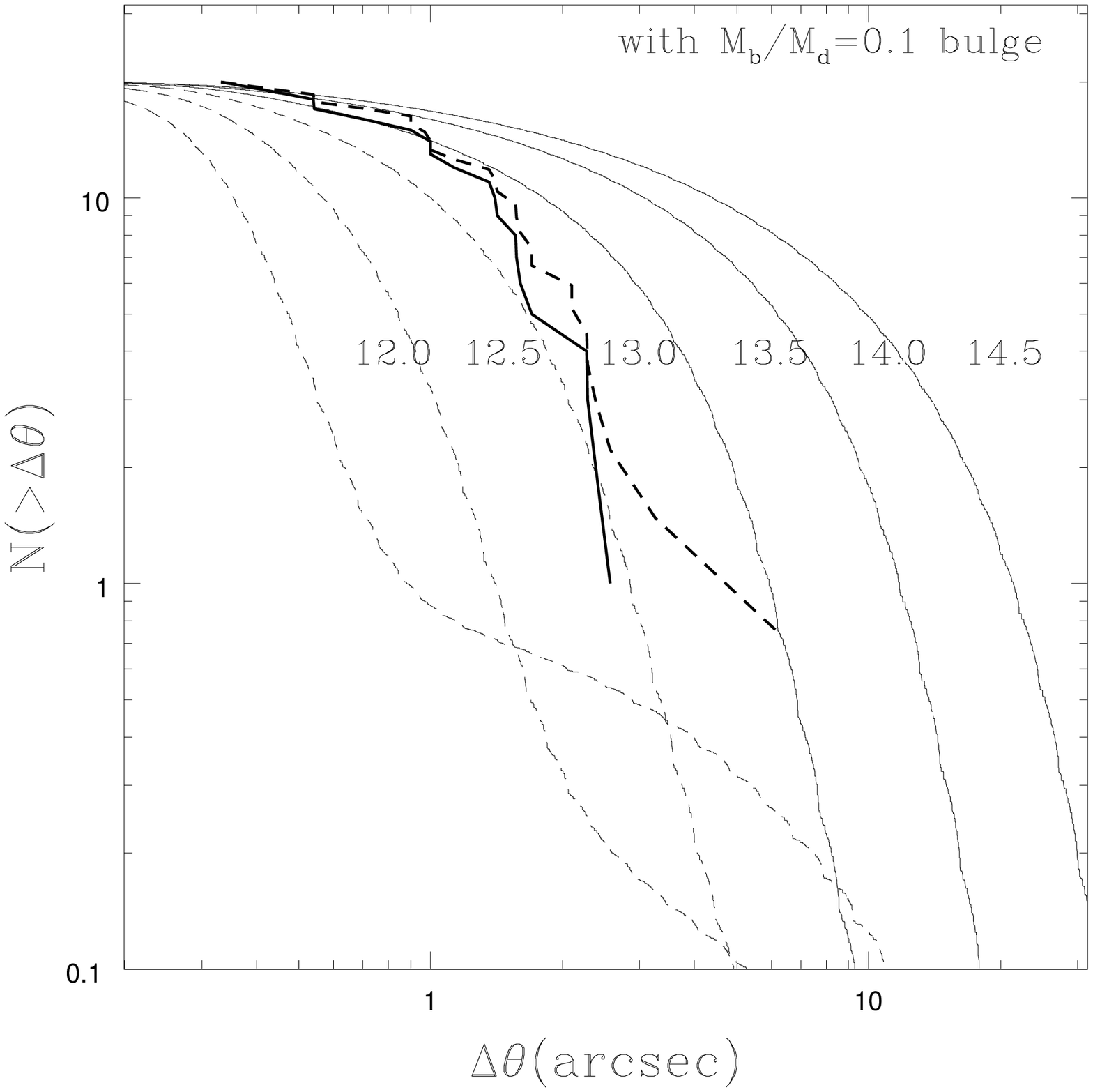}}
\end{center}
\caption{Predicted separation distributions with a cooling scale for models
with a $m_b/m_d=0.1$ bulge (right) and without a bulge (left).
The dashed curves show the distributions for $M_c=10^{12}M_\odot$,
$3\times 10^{12} M_\odot$ and $10^{13} M_\odot$ while the solid curves show
the distributions for $3\times 10^{13} M_\odot$ and $10^{14} M_\odot$
$3\times 10^{14} M_\odot$ and $10^{15} M_\odot$.  The curves are labeled
by $\log M_c$.  The heavy solid (dashed) curve shows the observed distribution
of the CLASS (radio-selected) lenses. }
\label{fig:sepdist}
\end{figure}

\begin{figure}
\begin{center}
\leavevmode
\epsfxsize=6in\epsfbox{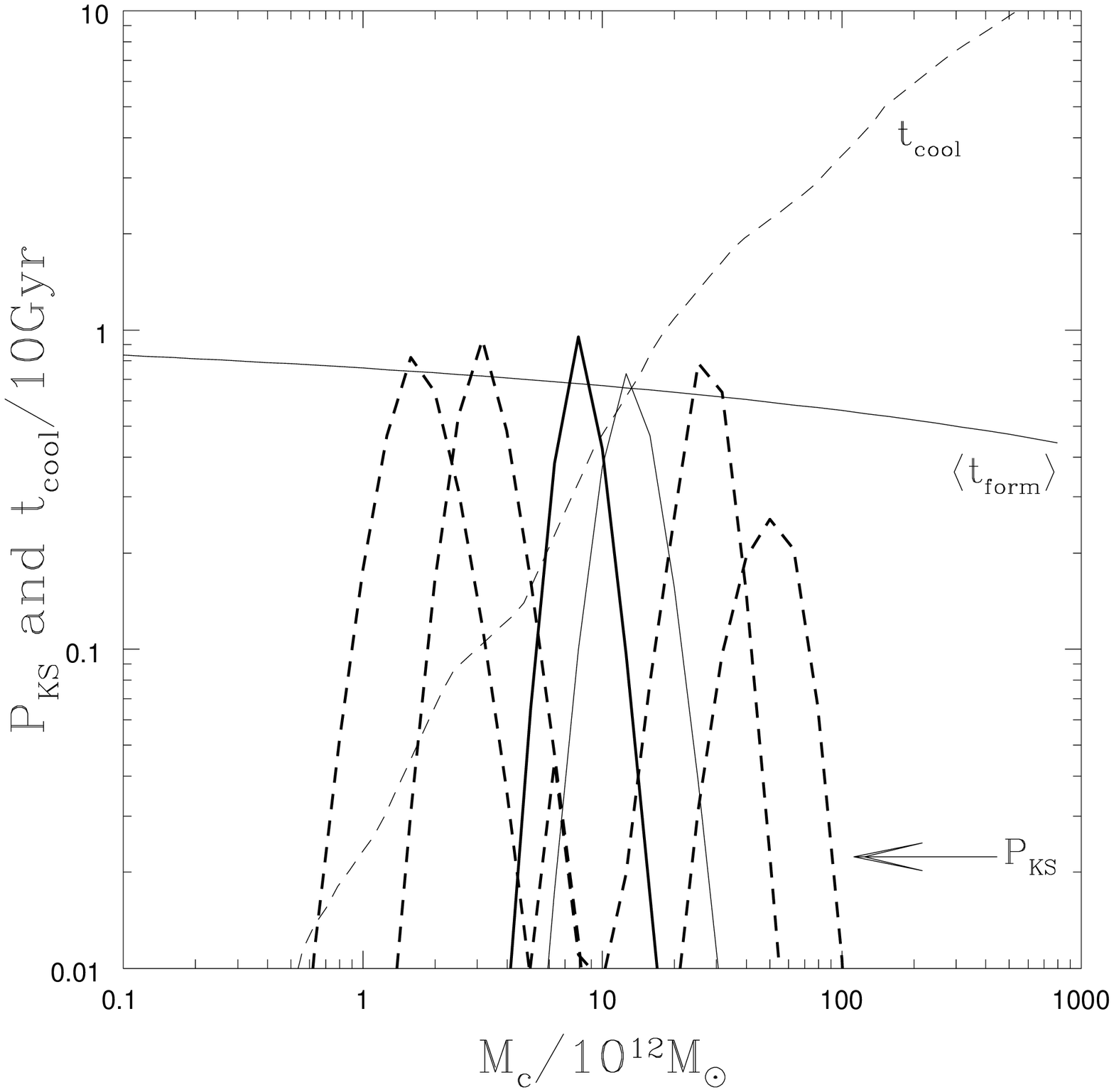}
\end{center}
\caption{The Kolmogorov-Smirnov probability, $P_{KS}$, of fitting the
observed separation distribution of CLASS lenses as a function of the
cooling mass scale $M_c$. The heavy (light) solid curves indicated by
the arrow show the K-S probability for models with $m_b+m_d=0.05$
without (with) a $m_b/m_d=0.10$ bulge.  The heavy dashed curves show
the K-S probabilities for models with lower ($m_b+m_d=0.01$ and $0.02$)
or higher ($m_b+m_d=0.10$ and $0.20$) baryon fractions where 
the optimal cooling mass decreases as the baryon fraction
rises.  The light dashed curves show the cooling time in units of 
$10$~Gyr for the radii enclosing 50\% of the baryonic mass for the standard
model.
The light solid line shows the time since the average formation epoch
($\langle t_{\rm form}\rangle$) in units of $10$~Gyr assuming $h=0.67$.}
\label{fig:ksprob}
\end{figure}

The solution is to introduce an abrupt change in the structure of the
objects at the mass scale $M_c$ dividing galaxies and clusters.
Porciani \& Madau (\cite{PorMad}) used SIS models normalized by the local
properties of galaxies below $M_c$ and NFW models above $M_c$ and could fit
the observed separation distribution given a mass scale of
$M_c=3\times10^{13}M_\odot$.
We use the adiabatically compressed models below $M_c$ and the NFW models
above $M_c$.
In our models, $M_c$ is simply the mass at which 50\% of halos have cooled
based on the probability distribution defined by Eq.~(\ref{eqn:pcool}).
The compressed halos have far higher cross sections per unit mass than the
original NFW halos (see \S\ref{sec:cool}), leading to an increase in the
fraction of small separation lenses.
Fig.~\ref{fig:sepdist} shows the predicted distributions as a function of
$M_c$.

If the cooling mass scale is too large, $M_c\gtorder 3\times10^{14}M_\odot$,
then we find the distribution predicted by the adiabatically compressed
models without a cooling scale (Fig.~\ref{fig:sepdist}) and cannot match the
observations.
If the cooling mass scale is too small, $M_c\ltorder 10^{12}M_\odot$, then
we find the distribution predicted by the NFW models for large separations
combined with a sharp peak at small separations.
Only if the break is at a mass scale $M_c\simeq 10^{13}M_\odot$ can the models
reproduce the observed distribution.
Interestingly, cosmological hydrodynamic simulations also find that the cooled
baryon fraction reaches 50\% on mass scales near $M_c\sim 10^{13}M_\odot$
(e.g.~Pearce et al.~\cite{Pearce99}).

We can quantify the goodness of fit by using the Kolmogorov-Smirnov test
(e.g.~Press et al.~\cite{NumRec})
to compute the likelihood $P_{KS}$ that the model produces a separation
distribution consistent with the observations.
Fig.~\ref{fig:ksprob} shows $P_{KS}$ as a function of $M_c$ for a range
of models.
For a fixed baryonic fraction ($m_d+m_b$), the optimal value depends on the
assumed structure of the low mass halos.
For example, in our models $M_c$ increases from
$M_c\simeq 5\times 10^{12}M_\odot$ to $M_c\simeq 10^{13}M_\odot$ when we add
a $m_b/m_d=0.1$ bulge. 
By increasing the lensing efficiency of the low mass galaxies relative to the
high mass galaxies (by making them more supercritical lenses), the addition 
of the bulge drives the estimate of the cooling mass scale upwards.
An SIS model would further increase the relative lensing efficiency of the low
mass galaxies, which probably explains why Porciani \& Madau (\cite{PorMad})
found a still higher break mass: $M_c\simeq 3 \times 10^{13}M_\odot$.

The optimal mass scale depends on the baryon fraction, $m_d+m_b$.  With
less baryons, a halo of a given mass becomes a less efficient lens producing
smaller image separations, so $M_c$ must increase to keep the observed break
at a fixed image separation scale.
If the baryon fraction becomes too low to significantly compress the halos
($m_d+m_b \ltorder 0.01$), it becomes impossible to explain the observations.
The mass scale required to explain the observations depends exponentially on
the baryon fraction, with $\log_{10} M_c/M_\odot\sim 13.6-(m_d+m_b)/0.15$
for the models without a bulge.
These trends are also shown in Fig.~\ref{fig:ksprob}.  

\subsubsection{A Self-Consistent Model} \label{sec:coolmodel}

These models, and the earlier models by Keeton (\cite{Kee98}) and
Porciani \& Madau (\cite{PorMad}), introduced the break mass as an
{\it ad hoc\/} means of separating galaxies and clusters into separate
lens populations.
Physically this break mass is the cooling mass scale, which divides halos
in which the baryons have cooled from those in which they have not, and its
value should be predictable from the basic physics of cooling.

As a first step we computed the cooling times at $z=0$ for the radius
encompassing 50\% of the baryons in halos with
$m_d+m_b=0.05$, and compared it to the time elapsed since the average
formation time of the halos.
These time scales are superposed on Fig.~\ref{fig:ksprob}.
As expected, the cooling times are comparable to the time available for
cooling near the mass scales required to explain the distribution of image
separations.
Note, however, that agreement between the cooling mass scale and the break
mass required to explain the lenses depends strongly on the baryon fraction
$m_d+m_b$.
The two scales agree for our fiducial model with $m_d+m_b=0.05$.
For higher baryon fractions the break mass required to explain the lenses is 
well below the cooling mass scale, and for lower baryon fractions it is well
above the cooling mass scale.
Fig.~\ref{fig:ksprob} underestimates the problem because the cooling time
also depends on the baryon fraction as
$\tau_{\rm cool}\propto\rho_{\rm gas}^{-1}$.
The cooling time drops when we increase the baryon fraction, which will
exacerbate the discrepancies between the cooling mass and the break mass
needed to explain the lenses.  
Thus, cooling physics makes the baryonic mass fraction, $m_d+m_b$, of the cooled
halos the key parameter relevant to determining the distribution of image
separations.
    
\begin{figure}
\begin{center}
\leavevmode
\centerline{\epsfxsize=3.5in\epsfbox{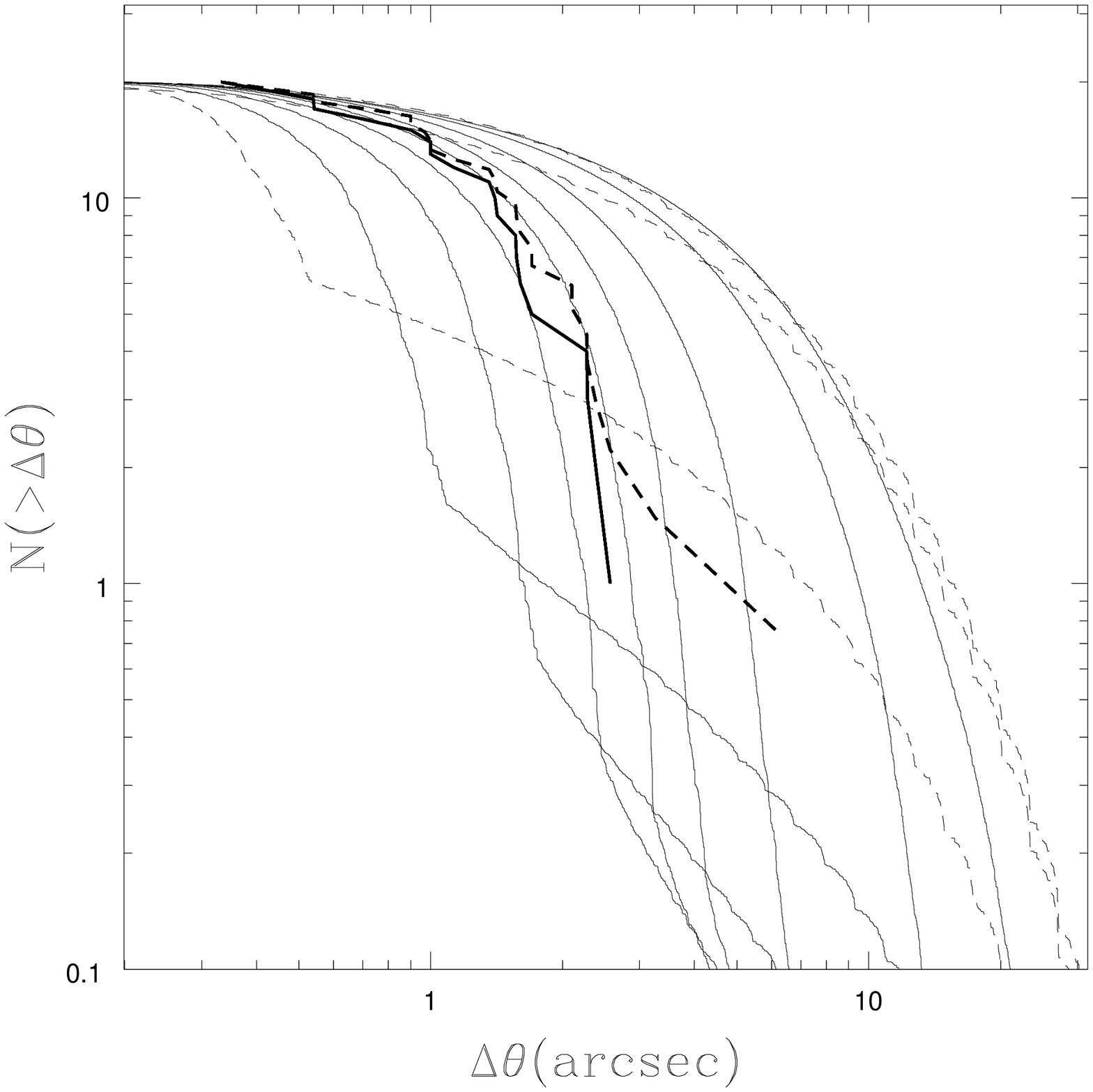}\epsfxsize=3.5in\epsfbox{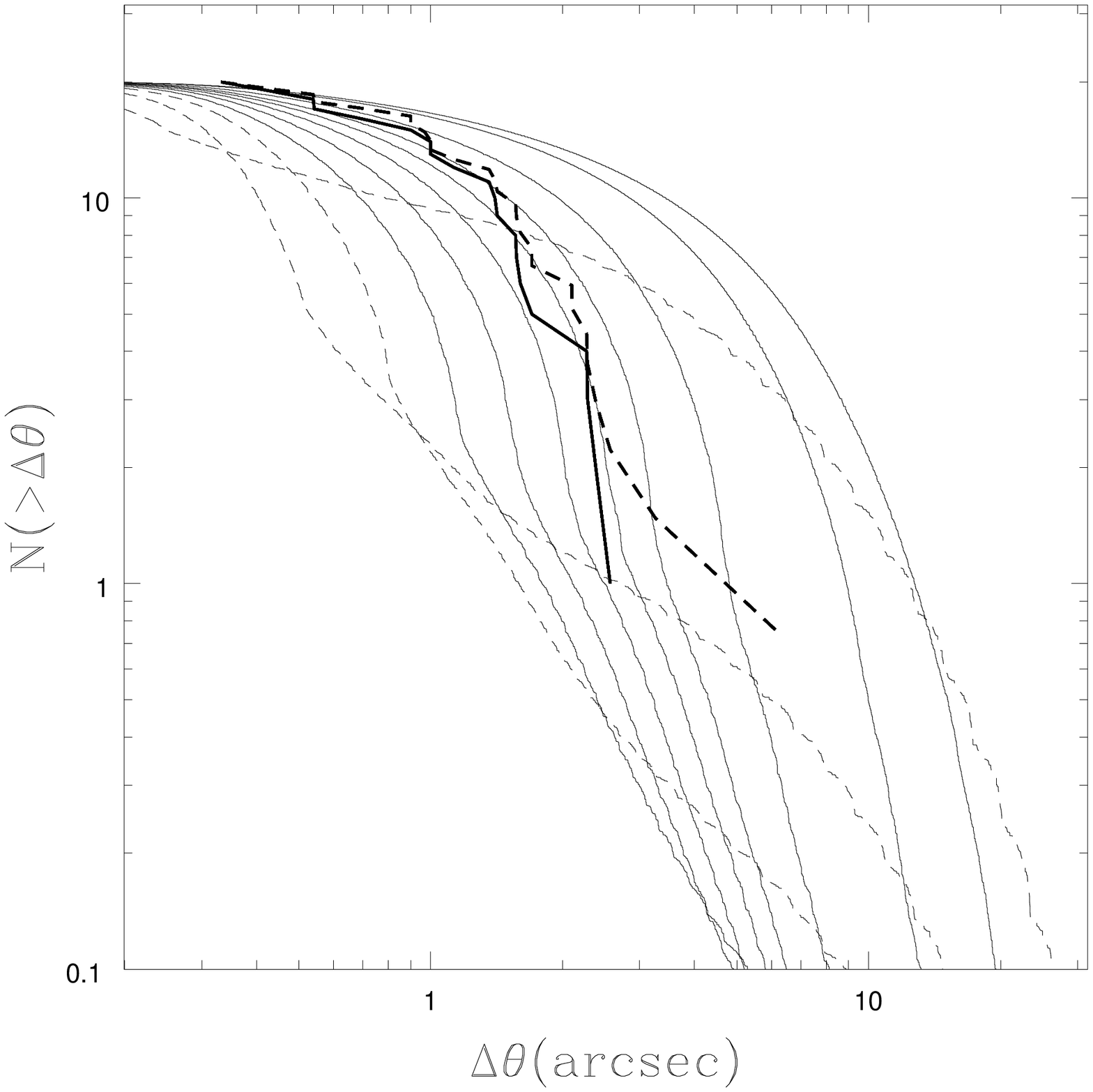}}
\end{center}
\caption{Image separation distributions as a function of
$\Omega_{b,{\rm cool}}$
using models with a $m_b/m_d=0.1$ bulge (right) or without a bulge (left).
The dashed curves show the distributions for $\Omega_{b,{\rm cool}}=0.003$,
$0.006$, and $0.009$ (from right to left at large separation), and the
solid curves show the distributions for $\Omega_{b,{\rm cool}}=0.0012$,
$0.015$, $0.018$, $0.021$, $0.024$, $0.030$, $0.045$ and $0.060$ 
(from left to right at large separation). The
heavy solid (dashed) show the observed separation distribution of the CLASS
(all radio) lenses.}
\label{fig:newsepdist}
\end{figure}

\begin{figure}
\begin{center}
\leavevmode
\epsfxsize=6in\epsfbox{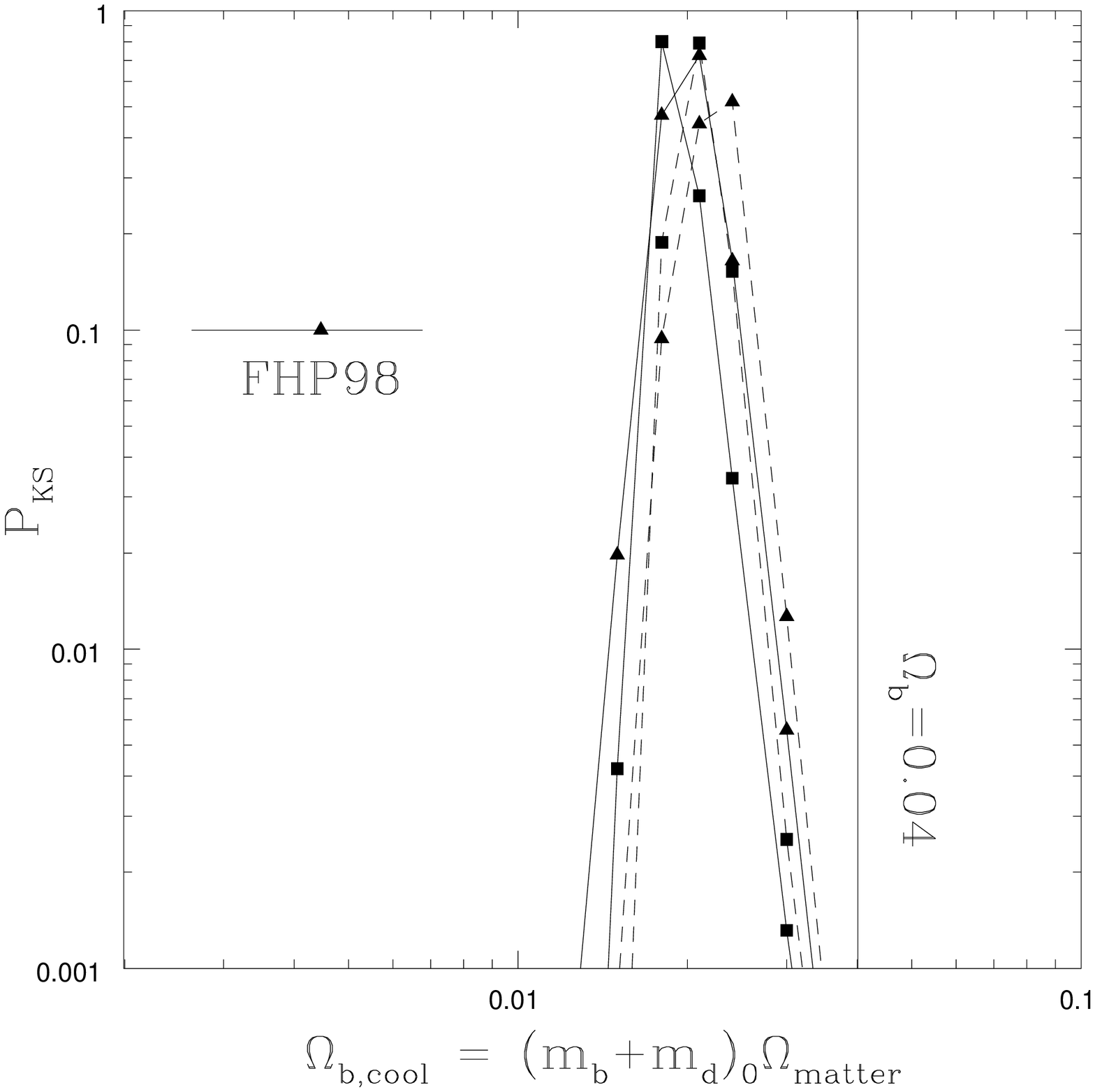}
\end{center}
\caption{Kolmogorov-Smirnov test probability of fitting the separation
distribution of CLASS lenses as a function of $\Omega_{b,{\rm cool}}$.
The squares (triangles) indicate models with no bulge (with a $m_b/m_d=0.1$
bulge), and the solid (dashed) lines correspond to fitting the CLASS lenses
(all radio lenses).  The point with horizontal error bar is the estimate
by Fukugita, Hogan \& Peebles~(\protect\cite{Fuk98}) for the cold baryon
(stars, remnants, cold gas) content of galaxies.  The vertical line marks
the total baryon content in the concordance model. }
\label{fig:newksprob}
\end{figure}

In our final models the only parameter is the halo baryon fraction
$(m_d+m_b)_0$ which should equal $\Omega_b/\Omega_m$ for halos which are 
a fair sample of the universe.  The cooled baryon fraction is then 
determined from our simple model of the cooling physics from \S2.3.
Figure~\ref{fig:newsepdist} shows the separation distributions computed 
using our cooling model as a function $\Omega_{b,{\rm cool}}$, where the
structure of each halo is set by the adiabatically compressed models with
a cooled baryon fraction of $m_b+m_d=f_{\rm cool}(M,z)(m_b+m_d)_0$ and
$(m_b+m_d)_0=\Omega_{b,{\rm cool}}/\Omega_m$.
The qualitative behavior of the models is similar to the more phenomenological
models based on the cooling mass scale $M_c$.
Low $\Omega_{b,{\rm cool}}$ models have difficulty cooling, making them
equivalent to models with a low cooling mass scale.
High $\Omega_{b,{\rm cool}}$ models cool easily, making them equivalent to
models with a high cooling mass scale.
Models with $0.015 \ltorder \Omega_{b,{\rm cool}} \ltorder 0.025$ agree
with the observed separation distributions independent of which data we fit
(CLASS lenses or all radio lenses) or the assumed density structure
(with or without a $m_b/m_d=0.1$ bulge).  The likelihoods as a function of 
$\Omega_{b,{\rm cool}}$ are shown in Fig.~\ref{fig:newksprob}.  The results
are insensitive to modest errors in the cooling function, as raising and
lowering the cooling curve by factors of two only affects the estimated
baryon fraction by 20\%.
With the introduction of the cooling physics, there is no trivial 
scaling of the results for the value of the Hubble constant.

While the preferred range is less than the total baryon density
$\Omega_b=0.04$ in the input cosmology, it significantly exceeds
the estimates of $0.0045 \ltorder \Omega_{b,{\rm cool}} \ltorder 0.0068$
for the cool baryon fraction (stars, cold gas and stellar remnants) in 
local galaxies by Fukugita, Hogan \& Peebles~(\cite{Fuk98}).
This discrepancy could have two explanations.
First, it could be a problem in our models.
The adiabatic compression models are crude approximations for the
transformation of the dark matter halos by the baryons.
While the estimates of $\Omega_{b,{\rm cool}}$ appear to be insensitive to
changes in our assumptions, we know that the observed lens population is
dominated by early-type galaxies whose baryonic density structure is very
different from the assumptions used in our models.
It is difficult to adequately address this possibility, since it is currently
impossible to compute the final structure of a galaxy starting from the initial
halo properties.
Second, the accounting for the baryons in galactic halos may be incorrect.
The Fukugita et al.~(\cite{Fuk98}) accounting for the bayrons in galaxies
included only cold gas components, neglecting hot ($10^6$~K) and warm,
ionized ($10^4$--$10^5$~K) components.
While hot gas cannot contribute the adiabatic compression of the halo, the
warm components are both difficult to detect and contribute to the
compression.
Perhaps our best route towards understanding the detailed physical
modifications of the dark matter halos by baryonic processes in hierarchical
models is through simulations, but at present these have proven extremely
difficult (see e.g.~Navarro \& Steinmetz~\cite{NavSte}).

\begin{figure}
\begin{center}
\leavevmode
\epsfxsize=6in
\epsfbox{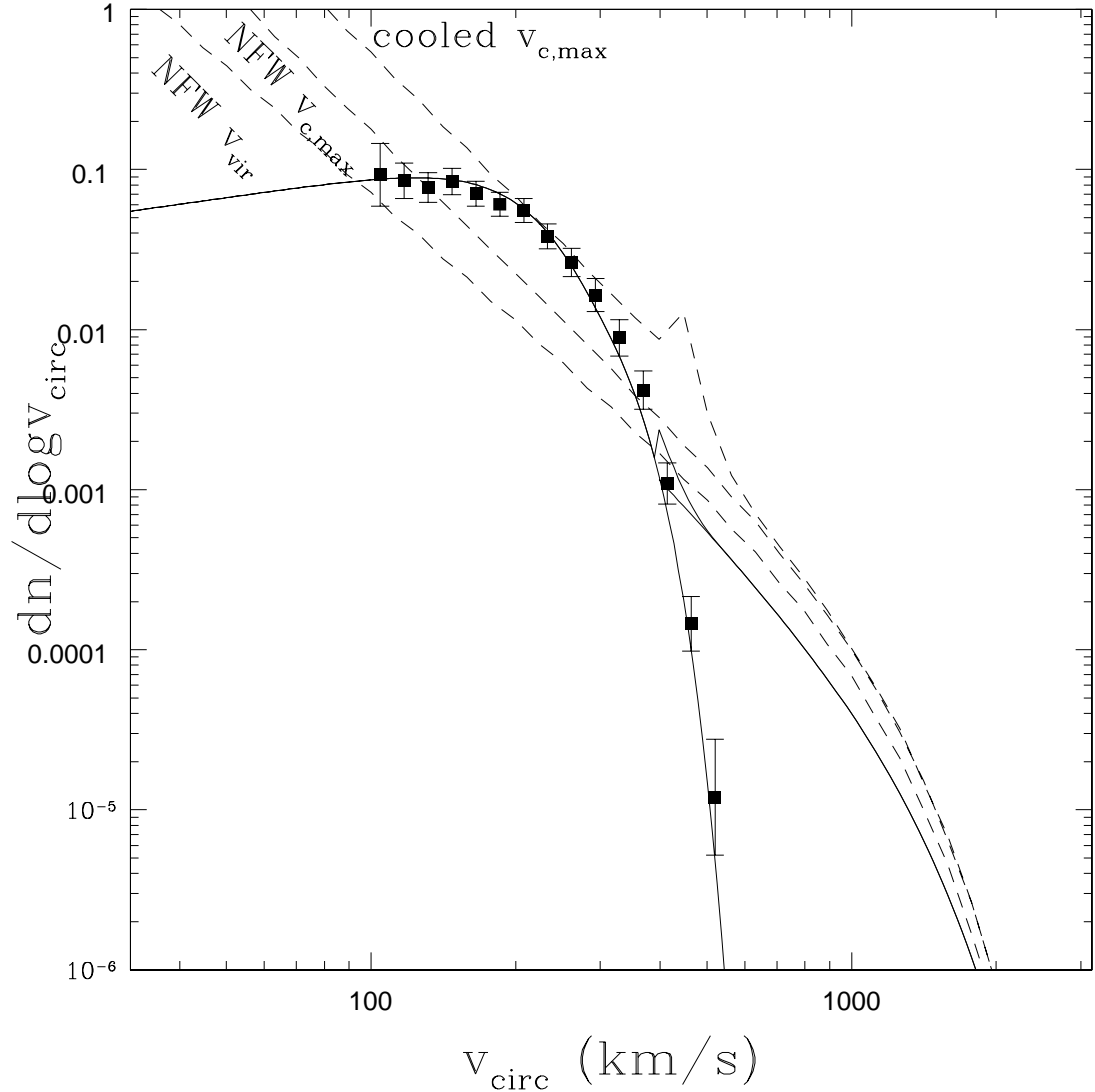}
\end{center}
\caption{The velocity function
$dn/d\log v_c = (dn/d\log M)|d\log M/d\log v_c|$.
The solid curves show the local velocity function of galaxies (low $v_{circ}$)
and clusters (high $v_c$) and their sum.
The points are the non-parametric velocity function of galaxies.  From bottom
to top, the dashed curves show the velocity functions derived using $dn/dM$
and the NFW virial velocity (labeled NFW $v_{\rm vir}$), the peak circular
velocity of the NFW rotation curve (labeled NFW $v_{c,{\rm max}}$) and the
peak circular velocity of the adiabatically compressed model (labeled cooled
$v_{c,{\rm max}}$).  We used the $\Omega_{b,{\rm cool}}=0.018$ model with no
bulge.}
\label{fig:veldist1}
\end{figure}

\begin{figure}
\begin{center}
\leavevmode
\epsfxsize=6.0in\epsfbox{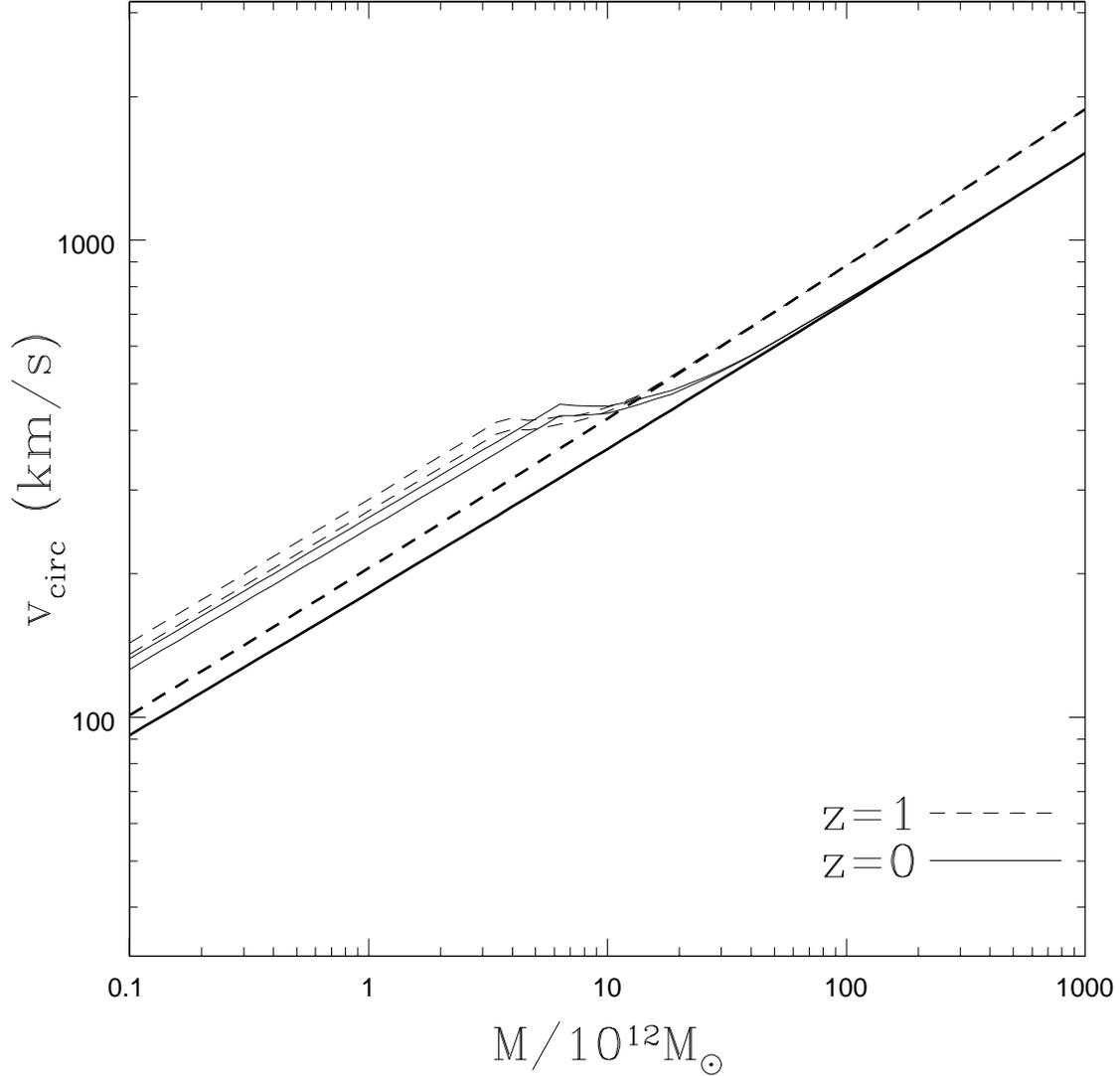}
\end{center}
\caption{The global relation between mass and circular velocity at redshifts
zero (solid) and unity (dashed).  The heavy curves show the peak circular
velocity of the NFW model.  The light curves show the peak circular velocity
including the baryonic cooling and adiabatic compression from the
$\Omega_{b,{\rm cool}}=0.018$ model.  The upper light curve is
the model with no bulge component ($m_b/m_d=0$) and the lower light
curve is the model with a bulge ($m_b/m_d=0.1$).  The bulge slightly
reduces the peak rotation velocity (because it increases the angular
momentum per unit mass of the disk) while making the rotation curve
flatter (see \S\protect\ref{sec:cool}). }
\label{fig:veldist0}
\end{figure}

\begin{figure}
\begin{center}
\leavevmode
\epsfxsize=6.0in\epsfbox{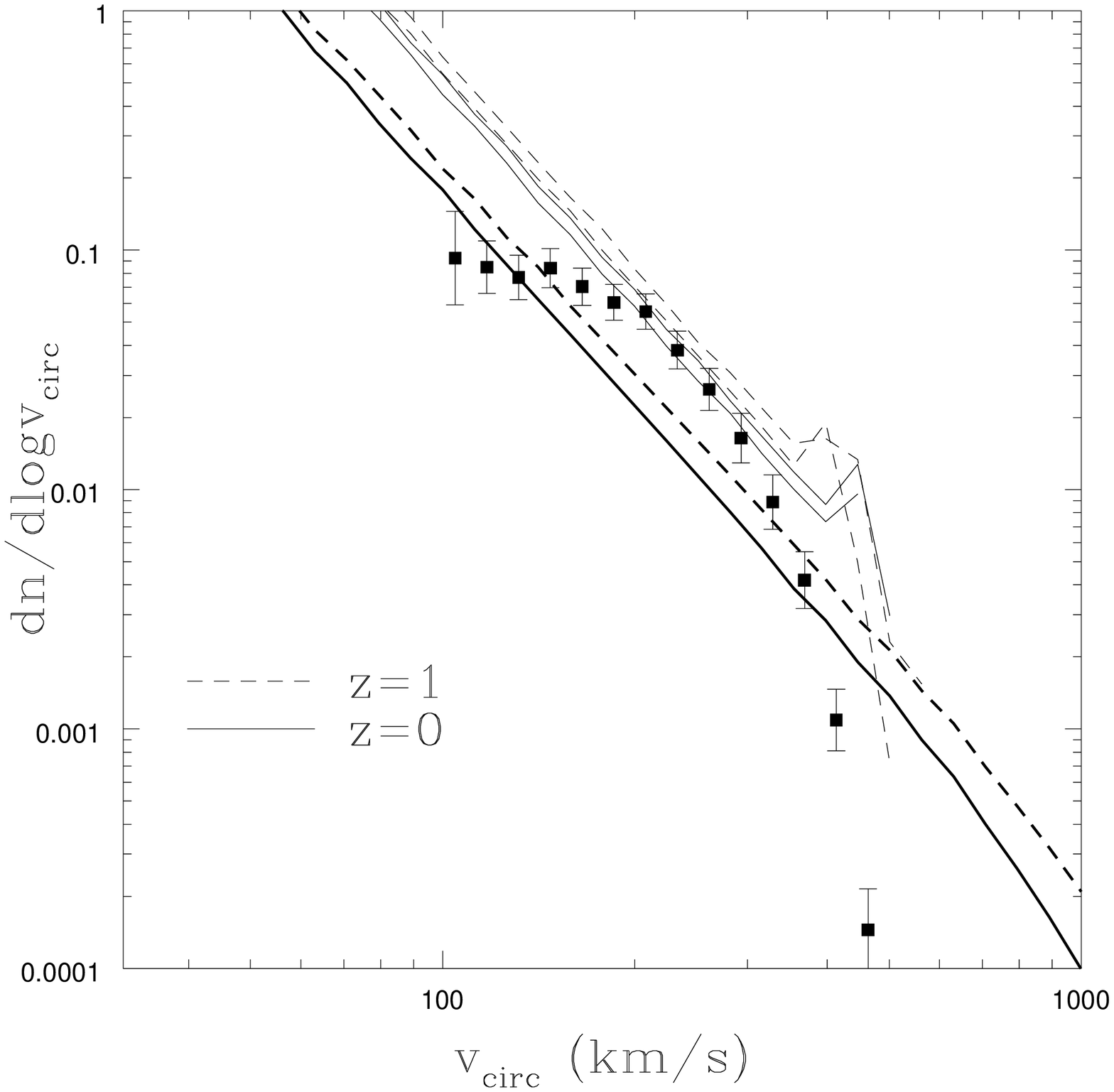}
\end{center}
\caption{The velocity function at redshift zero (solid) and unity (dashed).
The heavy curves show the results for the dark matter model.  The light 
curves show velocity functions for halos in which at least $1/3$ of the
baryons have cooled in the $\Omega_{b,{\rm cool}}=0.018$ model. The upper 
light curve is the model with no bulge component ($m_b/m_d=0$) and the 
lower light curve is the model with a bulge ($m_b/m_d=0.1$).  Superposed
on the distributions are the points from the velocity function of
galaxies at $z=0$ (see text).}  
\label{fig:veldist2}
\end{figure}

\section{The Distribution of Circular Velocities} \label{sec:velocity}

Given our success in fitting the distributions of gravitational lens image
separations, it is natural to ask whether our model agrees with estimates
of the local velocity function.
Global estimates of the local (circular) velocity function, $dn/d\log v_c$,
are difficult because galaxies and clusters have very different dynamical
properties and require different observational methods.
We first construct a rough estimate of the local velocity function including
both galaxies and clusters, and then compare it to our best self-consistent
models.
We stress that most of the pieces in our model have been addressed before by
other authors
(e.g.~Turner, Ostriker \& Gott~\cite{TOG84};
Cole \& Kaiser~\cite{ColKai};
Fukugita \& Turner~\cite{FT91};
Shimasaku~\cite{Shim93};
Kochanek~\cite{Koc95},~\cite{Koc96};
Gonzalez et al.~\cite{GWBKP};
Sigad et al.~\cite{SKBKKPD})
here we attempt to focus on the combination of cluster and galaxy scales to
emphasize the effect of cooling in producing a change in the velocity function.

We will use the velocity function for local galaxies derived by
Pahre, Kochanek \& Falco~(\cite{Pahre00}), which has significantly reduced
systematic uncertainties compared to earlier estimates.
The velocity function for groups and clusters is notoriously difficult to
estimate.
We will combine the X-ray temperature function estimate by
Blanchard et al.~(\cite{Blan00}) with the average X-ray relation between
the X-ray temperature and the galaxy velocity dispersion derived by
Wu et al.~(\cite{Wu99}, eqn. (14)) to estimate the velocity function for groups and
clusters.  This velocity function estimate is nearly identical to a simple
thermodynamic conversion using $v_c = (2 kT/\mu m_p)^{1/2}$ to estimate
the circular velocity.
There are non-trivial systematic uncertainties in our final result because of
the problems in uniformly relating late-type galaxy rotation curves,
early-type galaxy velocity dispersions and X-ray temperatures to a common
circular velocity scale.
While the upper velocity limits of galaxies are unambiguous, the lower
velocity limits for groups are unknown because of the difficulty in
unambiguously recognizing and counting low mass groups.
For simplicity we simply truncate the cluster contribution at temperatures
below 0.5~keV, which introduces a small kink in the distribution.
Fig.~\ref{fig:veldist1} shows the estimated velocity function, which
has a prominent break near $v_c \simeq 400$~km~s$^{-1}$ at the boundary
between the high-mass galaxies and the low-mass groups.  This break
in the velocity function is the local signature of the cooling mass
scale, and its amplitude is far larger than any systematic uncertainties
in our construction of the velocity function.

We can convert the halo mass function into the velocity function given
a model relating the mass $M$ to the circular velocity $v_c$.  Ignoring
the baryons we could use either the virial velocity of the NFW halo,
$v_{\rm vir}=(GM_{\rm vir}/r_{\rm vir})^{1/2}$ or the peak circular velocity
$v_{c,{\rm max}}$ of the rotation curve of the halo.
A variable transformation,
$dn/d\log v_c = (dn/d\log M)|d\log M/d\log v_c|$ relates the mass 
function to the velocity function.  Fig.~\ref{fig:veldist1} shows
the predictions for the velocity function, and Fig.~\ref{fig:veldist0}
shows the relationship between $M$ and $v_c$ for the two cases.  As
expected for a model normalized to match the abundance of rich clusters,
the dark matter model produces a reasonable match to the velocity
function of clusters given the uncertainties in the construction of the
velocity function.  The dark matter model catastrophically fails to 
match the velocity function of galaxies, grossly underpredicting the
density for $v_c \sim 200$~km~s$^{-1}$ and grossly overpredicting it
for $v_c \ltorder 100$~km~s$^{-1}$.  These discrepancies are created
by the effects of the baryons in distorting the halo structure
(see also Gonzalez et al.~\cite{GWBKP}).  

The break in the velocity function should be reproduced by our model
for the distribution of image separations.  Fig.~\ref{fig:veldist0}
shows the relationship between $v_c$ and $M$ for the model from
\S\ref{sec:cool} with $\Omega_{b,{\rm cool}}=0.018$ which provided a good
fit to the distribution of lens separations.  For large masses, where the 
baryons have not cooled, the peak circular velocity matches that
of the dark matter models.  For small masses, all the baryons have
cooled and the circular velocity curve is shifted upwards.  Near
$M \simeq 10^{13}M_\odot$ there is a break in the curve between
the uncompressed and fully compressed slopes.  The model velocity
function (Fig.~\ref{fig:veldist1}) now has a break at the same velocity 
scale as the observed velocity function.  The adiabatic compression 
shifts the more numerous low-mass halos to a higher circular velocity, 
bringing the density of halos with $v_c \sim 200$~km~s$^{-1}$ close to
that observed.
The models still overpredict the density of very low mass halos,
$v_c \ltorder 100$~km~s$^{-1}$ probably because star formation reheats
and/or disrupts low mass halos (see Gonzalez et al.~\cite{GWBKP}).
These low-mass halos had no impact on the distribution of lensed image
separations because the $\sim v_c^4$ scaling of the lensing cross sections
and the finite angular resolution of the surveys
($\Delta\theta \simeq 0\farcs25$)
makes the lens surveys insensitive to these mass scales.
To fix this problem we would need to adopt a similar scheme to that used in
the semi-analytic models (see, e.g., SA) to fix the overabundance of low-luminosity
halos: appeal to feedback to suppress star formation in these ``galaxies''.
 
While the break in the velocity function consistent with the distribution
of lens separations is located on the right velocity scale, the agreement
is not perfect, principally because our model produces a peak in the velocity 
function at approximately $v_c\simeq 500$~km~s$^{-1}$.  No similar peak 
is seen in the distribution of image separations because the separation
distribution is smoothed by the redshift distribution of the lenses. 
The peak is created (mathematically) by the region near $M=10^{13}M_\odot$
where the $v_c$-$M$ relation is flat.  The flat region of the 
the $v_c$-$M$ relation is produced by two problems in our model.  The first is the
``over-cooling'' problem common to many semi-analytic models (see SA) 
including the Cole et al.~(\cite{Cole00}) model we used.  The cooled mass and
the resulting star formation predicted for halos on group mass scales
are too large, leading to super-luminous galaxies which are not 
observed.  More complicated models can significantly reduce the problem
(see Cole et al.~\cite{Cole00}).  The second problem is that we
assumed a deterministic relationship between mass and circular velocity
by using an average formation time rather than a distribution of formation
times for each halo.  More complicated models which use a distribution
of formation times
(e.g.~Cole et al.~\cite{Cole00}; Newman \& Davis~\cite{NewDav})
would smear out the feature.

Newman \& Davis~(\cite{NewDav}) proposed using the evolution of the velocity
function of galaxies as a probe of the cosmological model.
Our experience here suggests that uncertainties in baryonic physics relating
the observed galaxy velocities to those of the ``underlying'' dark matter
halo could impact this proposal.
Fig.~\ref{fig:veldist0} shows the $v_c$-$M$ relation both today and at
redshift unity.
At fixed halo mass, the circular velocity of the dark matter models increases
with look back time because the average density of the halos increases
(Navarro, Frenk \& White~\cite{NFW,NFW2}).
With the inclusion of the baryons, the evolution of the $v_c$-$M$ relation is
considerably more complicated.  First, because the time available for
cooling is significantly less at redshift unity than today, the cooling
mass scale evolves with redshift.  In these models the cooling mass 
scale today is $M\simeq 6 \times 10^{12}M_\odot$, while at redshift
unity it is $M \simeq 3 \times 10^{12}M_\odot$, based on the point where
the $v_c$-$M$ relation begins to return to  the dark matter relation.  Because
the cooling mass determines the mass of the most massive and luminous galaxies,
magnitude limited studies of galaxies will be very sensitive to the evolution
of the cooling scale because they are dominated by galaxies near the 
characteristic luminosity created by the existence of a cooling mass scale.
Second, at lower masses where we might avoid problems due to the evolution
of the cooling mass scale, the evolution in the circular velocity at fixed
mass differs between the dark matter and adiabatically compressed models.
The adiabatic compression modestly reduces the change in the circular velocity
at fixed mass compared to the dark matter models.  Third, the relationship
between mass and circular velocity depends on the detailed of the distribution 
of the baryons.  The models with a bulge have somewhat lower peak rotation 
velocities than the models without because of our choice for distributing 
the baryonic angular momentum
(see \S\ref{sec:adiabatic} and \S\ref{sec:cool}).  

Fig.~\ref{fig:veldist2} shows the evolution of the velocity function based
on these different $v_c$-$M$ relations.  At fixed peak circular velocity,
the number density of halos increases between today and redshift unity by
about $25\%$.
We also show the evolution based on the peak circular velocity distribution
of halos in which at least $1/3$ of the baryons have cooled.
The number density at fixed circular velocity still increases, but the ratio
is slightly smaller.  We see that the choice of density model has reasonably
strong effects, and the behavior of the velocities near the cooling scale
becomes more complicated.  In fact the differences between the models with
and without a 10\% bulge are equal to the differences created by evolution.
This means that the details of the baryonic mass distribution, and the
evolution of the baryonic mass distribution are at least as important as
cosmology to the evolution of the velocity function of galaxies.
For example, in most models of galaxy formation, the distribution of bulge
parameters is itself an evolving quantity, if nothing else because the
relative numbers of late-type disk galaxies and early-type, bulge-dominated
galaxies evolves (see SA; for observational evidence, see,
e.g.~van Dokkum \& Franx~\cite{DokFra}).
We have also superposed the Pahre et al.~(\cite{Pahre00}) estimate of the
velocity function to emphasize that our theory, even with the baryons, shows
far larger differences in shape from the model predictions than the magnitude
of the evolutionary effects.
For these reasons we believe that the evolution of the velocity function with
redshift will most likely provide crucial information about the process of
galaxy formation, before it can be used to discriminate between different
models for the evolution of the expansion rate.

\section{Conclusions} \label{sec:conclusions}

The mass and velocity distributions of virialized objects include a feature,
the baryonic cooling scale $M_c$, which divides halos which host galaxies
from those which host groups and clusters.  The cooling not only produces 
differences in morphology, but also alters the density distributions and 
the dynamical structure of the halos.  This significantly influences the
relationship between observable kinematic properties of galaxies and the
properties of their ``primordial'' parent halo.
These baryonic effects can go a long way towards reconciling theoretical
predictions for the distribution of lensing separations with observations,
and explaining some of the features of the galaxy and cluster velocity
function.

The baryonic compression which helps support the rotation curves of galaxies
also leads to an enormous increase in the cross section for the halos to be
gravitational lenses, converting shallow NFW density profiles into profiles
which begin to resemble the steep singular isothermal spheres preferred as
models for gravitational lenses (Cohn et al.~\cite{Cohn01}).  
The sharp increase in the lens cross section produced by cooling the baryons
leads us to expect a ``break'' in the separation distribution of gravitational
lenses at the separation scale corresponding to the cooling mass scale.  
The existence of such a break was first discussed by Keeton~(\cite{Kee98})
to explain why the lenses found in systematic surveys were always associated
with galaxies even when the galaxies were group or cluster members.
A model incorporating an {\it ad hoc\/} break was fit to the distribution
of lensing separations by Porciani \& Madau~(\cite{PorMad}) and used to
constrain the cooling scale to the range $10^{13}M_\odot$.

We have shown that a simple, self-consistent model based on adiabatically
cooled NFW profiles can fit the observed distribution of gravitational
lens separations.  In this model the lenses select a cooled baryon density
$0.015\la \Omega_{b,{\rm cool}}\la 0.025$ independent of the lens sample we
fit or our modeling of a galactic bulge.
While our model is self-consistent, in the sense that it uses calculable
properties of dark matter halos to convert from unmeasurable quantities
to observable quantities, it is clearly simplistic at best.
In particular we have neglected heating processes associated with star
formation, leading to cold baryon fractions in groups and clusters which
are too high and to an overestimate of the number density of low circular
velocity galaxies or small separation lenses.
It is gratifying that despite these shortcomings, the main features of the
observations can still be understood in terms of this simple physics.

When we predict the local distribution of halos in their observed circular
velocity using the same model that fit the distribution of image separations,
we find that our model also has a feature in the local velocity function at
the velocity scale dividing galaxies from clusters.  The comparison also 
makes it clear that additional physics (star formation, feedback and realistic
treatment of the differences between disks and bulges) and significant 
improvements in the cooling model are necessary to make the agreement 
any better than qualitative, particularly at the low $v_c$ end of the 
distribution (as has been noted by e.g.~Gonzalez et al.~\cite{GWBKP}).

The shortcomings of our model related to star formation and feedback could be
improved by using full semi-analytic models.
The density distribution on the other hand defies a simple treatment.
Our ``galaxies'' still have very large cores, even with the bulges, and would
likely predict observable central images, which are never seen in practice.
To make our models more realistic we would need to have density distributions
which are closer to isothermal spheres, which we would need to put in by hand
as done in the semi-analytic models or the earlier lens studies of
Keeton~(\cite{Kee98}) and Porciani \& Madau~(\cite{PorMad}).

Finally, when we explore the evolution of the velocity function of galaxies,
we find that the effects of the baryons are non-negligible.
It is only at extremely large radii that the ``velocity'' of a galaxy provides
a measure of the mass uncontaminated by baryonic processes.  On smaller scales
the measured quantity is related to the desired one by complicated and poorly
understood physics.
Further, the shape of the observed velocity function at low-$v_c$ suggests
that either observational selection effects cause a drastic underestimate of
low velocity systems or the efficiency of star formation is much lower in
these systems than their higher velocity counterparts (or both).
We do not currently understand how these effects would evolve with redshift.
In addition, the amount of baryons, the fraction of the baryons which cool and
its evolution, and the evolution in the average distribution of the baryons
(e.g.~bulge fractions) all produce changes in the velocity function that are
at least as important as the evolution in the underlying mass function.

\section*{Acknowledgements}

M.W.~would like to thank J. Newman and M. Davis for helpful conversations
on the velocity function and extended Press-Schechter theory and J. Cohn
and C. Keeton for a careful reading of the manuscript.
We thank R. Croft for discussions on the warm gas fraction in galaxies.
C.S.K. was supported by the Smithsonian Institution and NASA grant 
NAG5-8831.  M.W. was supported by a Sloan Fellowship and the National 
Science Foundation.

\appendix

\section{Simplifications to the Lens Calculations} \label{sec:simplify}

To estimate the lensing properties of the system we assume we view the
halo face on to the disk and then compute the mass $M(R)$ enclosed by the
cylindrical radius $R$.
By doing the calculations for a face-on disk we will underestimate the total
lensing cross section (see Keeton \& Kochanek~\cite{KeeKoc}).
The bending angle of the lens is then $\alpha_0 = 4 G M(<R)/c^2 R$ and the
lensed images are found as the solutions of
\begin{equation}
     u = { R \over D_{OL}} \pm \alpha_0(R) { D_{LS} \over D_{OS} }
       = x \pm \alpha(x)
\end{equation}
where $u$ is the angular source position, $x=R/D_{OL}$ is the angular
position of the image, $\alpha(x) =(D_{LS}/D_{OS})\alpha_0(D_{OL}x)$
and $D_{OL}$, $D_{LS}$ and $D_{OS}$ are angular diameter distances between
the Observer, Lens and Source (see Schneider, Ehlers \& Falco~\cite{Sch92}).

As is true for any circular lens, the tangential critical line (Einstein
ring) is at the solution $x_+$ of $1-\alpha(x_+)=0$ and the radial critical
line $x_-$ is at the solution of $1-\alpha'(x_-)=0$.
The radial caustic lies at $u_- = |x_- - \alpha(x_-)|$ and the multiple
image cross section is $\sigma = \pi u_-^2$.
When the source is on the radial caustic, there are two images on the critical
line at $x_-$ and an additional solution at $x_{\rm out}$ such that the mean
magnification produced by the lens in the multiply-imaged region is
$\langle M \rangle = (x_{\rm out}/u_-)^2$.

For a survey limited to sources brighter than flux $F_{\rm min}$, the
probability of finding a lens is
\begin{equation}
  p(F_{\rm min}) = \int D^2 dD \int d M { dn \over d M} \pi u_-^2
  B(F_{\rm min})
\label{eqn:plens}
\end{equation}
where $D^2 dD $ is the comoving volume element, $dn/dM$ is the redshift
and mass dependent halo density, $\pi u_-^2$ is multiple imaging cross
section, and $B(F_{\rm min})$ is the magnification bias factor.
We require the magnification bias term even when we consider only the
distribution of image separations because the average magnifications
produced by the the models differ.  We will use the Einstein ring
diameter ($2x_+$) to estimate the image separation $\Delta\theta$ 
produced by the lens, and we compute the integral distribution in
separation by appropriately adjusting the
limits of the integrals in Eq.~(\ref{eqn:plens}).

The uncertainties in the lensing properties of the halos are dominated by
the assumptions and problems in deriving the mass distributions.
These make estimates of the absolute lensing probabilities problematic.
Therefore we have focused only on the relative lensing probabilities,
particularly the distribution of the lenses in image separation.
Because the shapes of the potentials vary widely for the different lens
models, we need to include an estimate of the variations in the magnification
bias between the models rather than simply computing the multiple image cross
sections.
It is a generic feature of lens models that when the cross section is reduced
by rearrangements in the structure of the lens, the mean magnification rises.
Thus the resulting increase in the magnification bias makes the probability of
finding the lens change less rapidly than the cross sections (see, e.g. 
Kochanek~\cite{Koc96}).
We need a simple model which can rapidly estimate the effects of magnification
bias.
Fortunately, the magnification bias of the radio lens surveys is modest
because the slope of the luminosity function for radio sources fainter than
the survey flux limit is flat and the estimates will be insensitive to
simplifications in the calculation. 

We made two simplifications in the properties of the radio sources being
lensed.
First, we assumed the sources lay at a fixed redshift $z_s=2$ rather than
integrating over a distribution of lens redshifts.
Second, at that redshift we used an approximate version of the pure luminosity
evolution flat-spectrum radio luminosity function from
Dunlop \& Peacock (\cite{DunPea}).
For $x=P/P_c(z)$, Dunlop \& Peacock (\cite{DunPea}) used a luminosity function
of the form $dn/d\log x = d/(x^\alpha+x^\beta)$ with $\alpha=0.83$ and
$\beta=1.96$.
We approximated this form by a broken power-law, $\rho = d/x^\alpha$ for
$x < 1$ and $d/x^\beta$ for $x > 1$, to simplify the calculations of
magnification bias.
The approximate luminosity function differs from the original only near the
knee.
The break in the luminosity function lies at $\log P_c(z)=a_0+a_1 z+a_2 z^2$,
where $a_0=25.26$, $a_1=1.18$ and $a_2=-0.28$ and $P$ (in units of
W~Hz$^{-1}$~str$^{-1}$) is related to the flux (in Jy) by
$P=F D_0^2 (1+z)^{1+\alpha}$, $\alpha=0$ is the average spectral index assumed
for flat spectrum sources, and $D_0/(1+z)$ is the angular diameter distance in
an $\Omega_m=1$ cosmology.
The density factor $d$ will not enter our calculation, since we considered
only a fixed source redshift, but it is $d_0 (dV_0/dz)/(dV/dz)$ where $d_0$
is a constant density, $dV_0/dz$ is the comoving volume factor in an
$\Omega_m=1$ cosmology and $(dV/dz)$ is the comoving volume factor in the
cosmology we used for our calculation.
These corrections are required to convert the Dunlop \& Peacock (\cite{DunPea})
model from the $\Omega_m=1$ model in which it was derived to the cosmology we
are using in our calculation.

We then simplified the magnification bias calculation by assuming that the
magnification probability distribution was $P(>M)=(M_0/M)^2$ for $M \geq M_0$
where we could estimate the minimum magnification from the mean magnification
produced by the lens, $2 M_0 = \langle M\rangle$.
This approximation is exact for a SIS lens and the functional form is
asymptotically correct for any lens dominated by fold caustics.
By combining our approximate magnification distribution with our broken power
law model model for the luminosity function, we can analytically compute the
magnification bias for CLASS sample.

\end{document}